\documentstyle[prl,aps,epsfig,floats]{revtex}

\newcommand\beq{\begin{equation}}
\newcommand\eeq{\end{equation}}
\newcommand\bea{\begin{eqnarray}}
\newcommand\eea{\end{eqnarray}}
\newcommand\non{\nonumber}
\newcommand\noi{\noindent}

\begin{document}

\draft

\textheight=23.8cm
\twocolumn[\hsize\textwidth\columnwidth\hsize\csname@twocolumnfalse\endcsname

\title{\Large Renormalization group study of the conductances of
interacting quantum wire systems with different geometries}
\author{\bf Sourin Das$^1$, Sumathi Rao$^1$ and Diptiman Sen$^2$} 
\address{\it $^1$ Harish-Chandra Research Institute, Chhatnag Road, Jhusi,
Allahabad 211019, India \\
\it $^2$ Centre for Theoretical Studies, Indian Institute of Science, 
Bangalore 560012, India}

\date{\today}
\maketitle

\begin{abstract}
We examine the effect of interactions between the electrons on the 
Landauer-B\"uttiker conductances of some systems of quantum wires with 
different geometries. The systems include a long wire with a stub in the 
middle, a long wire containing a ring which can enclose a magnetic flux, and a
system of four long wires which are connected in the middle through a fifth 
wire. Each of the wires is taken to be a weakly interacting Tomonaga-Luttinger
liquid, and scattering matrices are introduced at all the junctions present in
the systems. Using a renormalization group method developed recently for 
studying the flow of scattering matrices for interacting systems in one 
dimension, we compute the conductances of these systems as functions of the 
temperature and the wire lengths. We present results for all three regimes of 
interest, namely, high, intermediate and low temperature. These correspond
respectively to the thermal coherence length being smaller than, comparable to
and larger than the smallest wire length in the different systems, i.e., the 
length of the stub or each arm of the ring or the fifth wire. The 
renormalization group procedure and the formulae used to compute the 
conductances are different in the three regimes. In particular, the
dimensionality of the scattering matrix effectively changes when the thermal
length becomes larger than the smallest wire length. We also present a 
phenomenologically motivated formalism for studying the conductances in the 
intermediate regime where there is only partial coherence. At low temperatures,
we study the line shapes of the conductances versus the energy of the electrons
near some of the resonances; the widths of the resonances are found to go to 
zero with decreasing temperature. Our results show that the Landauer-B\"uttiker
conductances of various systems of experimental interest depend on the 
temperature and lengths in a non-trivial way when interactions are taken into 
account. 
\end{abstract}
\vskip .6 true cm

\vskip.5pc
]
\vskip .6 true cm

\section{\bf Introduction}

The increasing sophistication in the fabrication of semiconductor 
heterostructures and carbon nanotubes in recent years have made it possible to
study electronic transport in different geometries. For instance, three-arm and
four-arm quantum wire systems have been fabricated by voltage-gate patterning
on the two-dimensional electron gas in GaAs heterojunctions 
\cite{timp,shepard}. Other systems of interest include Y-branched carbon 
nanotubes \cite{papa}, crossed carbon nanotubes \cite{kim1}, mesoscopic rings 
\cite{petrashov,casse}, and quantum wire systems with stubs \cite{debray}.
There have also been many theoretical studies of transport in systems
with various geometries \cite{buttiker1,jayannavar,shi,deo,kim2}.

Studies of ballistic 
transport in a quantum wire (QW) have led to a clear understanding of the 
important role played by both scattering of the electrons and the interactions 
between the electrons inside the QW \cite{kane,safi,maslov,lal1}. The 
scattering can occur either due to impurities inside the QW or at the contacts
lying between the QW and its reservoirs. A theoretical analysis using 
bosonization \cite{gogolin} and the renormalization group (RG) method 
typically shows that repulsive interactions between electrons tend to
increase the effective strength of the back-scattering as one goes to longer
length scales; experimentally, this leads to a power-law decrease in the 
conductance as the temperature is reduced or the wire length is increased 
\cite{tarucha}. Motivated by this understanding of the effects of interaction 
on scattering, there have been several studies of the interplay between the 
effects of interactions on one hand, and either a single junction between 
three of more QWs \cite{nayak,komnik,lal2,chen}, or more complicated 
geometries \cite{zhu,ben,sumathi,kim3} on the other. Using a RG technique 
introduced in Ref. \cite{yue}, the effects of a junction (which is 
characterized by an arbitrary scattering matrix $S$) has been studied in some 
detail \cite{lal2}. It is now natural to extend these studies to systems of 
QWs which are of experimental interest and which can have with more 
complicated geometries involving more than one junction.

In this paper, we will study the effect of interactions on the 
Landauer-B\"uttiker conductances of three systems of quantum wires with 
different geometries. These systems are shown in Figs. 1-3, and we will refer
to them as the stub, the ring and the four-wire system 
respectively. The stub system consists of two long wires, labeled as 1 
and 3, with a stub labeled as 2 being attached to the junction of 1 and 3.
The ring consists of two long wires, labeled as 1 and 3, between which 
there is a ring which can possibly enclose a magnetic flux; the two arms of 
the ring, labeled as 2 and 4, will be assumed to 
have the same length for convenience. The four-wire system consists of four 
long wires labeled as 1, 2, 3 and 4. The junction of 1 and 2 is connected 
to the junction of 3 and 4 by a fifth wire labeled as 5. 
The length of wire 2 in the stub system, the length of each of the arms
2 and 4 in the ring system, and the length of wire 5 in the four-wire system
will all be denoted by $L_S$. Each of the junctions present in the different 
systems is governed by a $3 \times 3$ scattering matrix $S$ which is unitary. 
We will assume that each of the wires in the various systems can be described 
as a one-channel weakly interacting Tomonaga-Luttinger liquid (TLL). For 
simplicity, we will ignore the spin of the electrons in this paper.

\begin{figure}[htb]
\begin{center}
\epsfig{figure=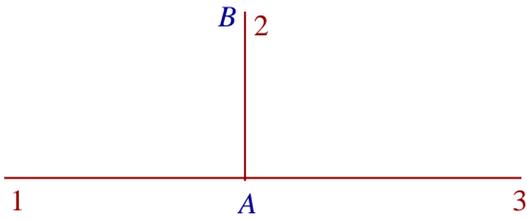,width=7cm}
\end{center}
\caption{The stub system, showing two long wires labeled as
1 and 3, and a stub labeled as 2. The lower end of the stub where three wires
meet and the upper end of the stub are denoted by $A$ and $B$ respectively.}
\end{figure}

\begin{figure}[htb]
\begin{center}
\epsfig{figure=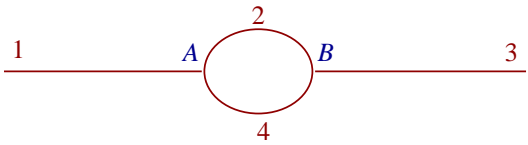,width=7cm}
\end{center}
\caption{The ring system, showing two long wires labeled as
1 and 3, the two arms of the ring labeled as 2 and 4, and two three-wire
junctions labeled as $A$ and $B$.}
\end{figure}

\begin{figure}[htb]
\begin{center}
\epsfig{figure=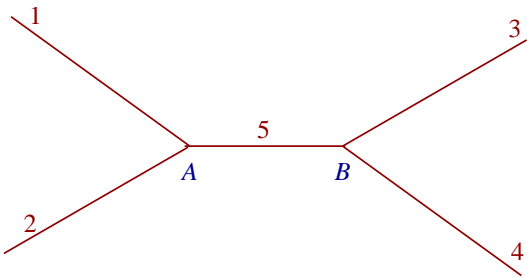,width=7cm}
\end{center}
\caption{The four-wire system, showing four long wires labeled as
1, 2, 3 and 4, a connecting wire in the middle labeled as 5, and two 
three-wire junctions labeled as $A$ and $B$.}
\end{figure}

In Sec. II, we will first summarize the RG method 
developed in Ref. \cite{lal2} for studying the flow of the $S$-matrix at a 
junction due to the interactions in the different wires connected to that 
junction. We will then describe our method for carrying out 
the RG analysis of the $S$-matrices at the various junctions of the different 
systems. In Sec. III, we will describe the procedure for computing the 
transmission probabilities (and conductances) of a system given the form of 
the $S$-matrices at all its junctions. It turns out that both the RG 
procedure and the route from the $S$-matrix to the conductances depend on the 
range of temperatures that one is considering. There is a length scale, 
called the thermal coherence length $L_T$, which governs the 
typical distance beyond which the phase of the electron wave function becomes
uncorrelated with its initial phase. The regimes of high, intermediate and low
temperatures are governed respectively by the condition that $L_T$ is much 
smaller than, comparable to or much larger than the length scale $L_S$ defined
above for the three systems; correspondingly, we have complete 
incoherence, partial coherence and complete coherence for the phase. The 
intermediate temperature range is the most difficult one to study, both for 
using the RG method and for computing the conductances. Based on some
earlier ideas \cite{buttiker2,mclennan}, we will describe a phenomenological
way of introducing partial coherence which will lead to expressions for the
transmission probabilities which interpolate smoothly between the 
coherent and incoherent expressions.

In Secs. IV-VI, we will apply the formalism outlined in the previous sections
to the stub, ring and four-wire systems respectively. In each case, the 
transmission probabilities at intermediate and low temperatures (i.e., the 
partially and completely coherent regimes) will be found to depend 
sensitively on the phase $\eta = e^{i2k_F L_S}$; here $k_F$ is the wave 
number of the electrons which are assumed to come into or leave the QW system 
with a momentum which is equal to the Fermi momentum in the reservoirs. In 
particular, certain values of $\eta$ can lead to resonances and anti-resonances
i.e., maxima and minima in the transmission probabilities. In the ring system,
there is another important phase which governs the possibility of resonance, 
namely, $e^{ie\phi_B /\hbar c}$, where $\phi_B$ is the magnetic flux enclosed 
by the ring, and $e$ and $c$ are the electron charge and the speed of light 
respectively. In each system, we will see how the conductances vary with the
temperature in a non-trivial way as a result of the interactions. This is 
the main point of our paper, namely, that interactions between the electrons
lead to certain power-laws in the temperature and length dependences of the 
conductances of experimentally realizable quantum wire systems.

\section{\bf Renormalization Group Method for Systems with Junctions}

In this section, we will first present the RG procedure developed in Ref.
\cite{lal2} for studying how the effect of a single junction
varies with the length scale. We will then describe how the RG method
has to be modified when a system has more than one junction.

A junction is a point where $N$ semi-infinite wires meet. Let us denote the
various wires by a label $i$, where $i=1,2,\cdots,N$. As we approach the 
junction, the incoming and outgoing one-electron wave functions on wire $i$ 
approach values which are denoted by $\psi_{Ii}$ and $\psi_{Oi}$ respectively;
we can write these more simply as two $N$-dimensional columns $\psi_I$
and $\psi_O$. The outgoing wave functions are related to the incoming ones
by a $N \times N$ scattering matrix,
\beq
\psi_O ~=~ S ~\psi_I ~.
\eeq
Current conservation at the junction implies that $S$ must be unitary. 
(If we want the junction to be invariant under time reversal, $S$ must also 
be symmetric). The diagonal entries of $S$ are the reflection amplitudes 
$r_{ii}$, while the off-diagonal entries are the transmission amplitudes 
$t_{ij}$ to go from wire $j$ to wire $i$. We will assume that the entries
of $S$ do not have any strong dependence on the energy of the electrons.

Let us now consider a simple model for interactions between the electrons, 
namely, a short-range density-density interaction of the form
\beq
H_{\rm int} ~=~ \frac{1}{2} ~\int \int dx dy ~\rho (x) ~V (x-y) ~\rho (y) ~,
\label{hint1}
\eeq
where $V(x)$ is a real function of $x$, and the density $\rho (x)$ is given in
terms of the second-quantized fermion field $\Psi (x)$ as $\rho = \Psi^\dagger
\Psi$. [The assumption of a short-ranged interaction is often made in the 
context of the TLL description of systems of interacting fermions in one 
dimension.] We define a parameter $g_2$ which is related to the Fourier 
transform of $V(x)$ as $g_2 = {\tilde V} (0)- {\tilde V} (2k_F)$. 
Different wires may have different values of this parameter which we will 
denote by $g_{2i}$. For later use, we define the dimensionless constants
\beq
\alpha_i ~=~ \frac{g_{2i}}{2\pi \hbar v_F} ~,
\label{ali}
\eeq
where we assume that the velocity 
\beq
v_F ~=~ \frac{\hbar k_F}{m}
\label{vf}
\eeq
is the same on all wires. In this work, we will be interested in the case in 
which the interactions are weak and repulsive, i.e., the parameters $\alpha_i$
are all positive and small.

We are now ready to present the RG equation for the matrix $S$ which was 
derived in Ref. \cite{lal2}. It will be useful to briefly discuss the 
derivation of the RG equation. A reflection from a junction, denoted by the 
amplitude $r_{ii}$ in wire $i$, leads 
to Friedel oscillations in the electron density in 
that wire. If $x$ denotes the distance of a point from the junction, the form 
of the oscillation at that point is given by the imaginary part of $r_{ii} 
e^{i2k_Fx} /(2\pi x)$. As a result of the interactions, an electron traveling
in that wire gets reflected from these oscillations. The amplitude of the 
reflection from the oscillations is proportional to $\alpha_i r_{ii} /2$ if 
the electron is reflected away from the junction, and to $\alpha_i 
r_{ii}^* /2$ if the electron is reflected towards the junction. These
reflections renormalize the bare $S$-matrix which characterizes the junction
at the microscopic length scale. The entries of $S$ therefore become functions
of the length scale $L$; we define the logarithm of the length scale as 
$l = {\rm ln} (L/d)$, where $d$ is a short-distance cutoff such as the 
average interparticle spacing. It is convenient to define a $N \times N$ 
diagonal matrix $M$ whose entries are given by
\beq
M_{ii} ~=~ \frac{1}{2} ~\alpha_i r_{ii} ~.
\eeq
Then the RG equation for $S$ is found to be \cite{lal2}
\beq
\frac{dS}{dl} ~=~ M ~-~ S M^\dagger S
\label{rg}
\eeq
to first order in the $\alpha_i$. (This equation is therefore perturbative in
the interaction strength).
One can verify from Eq. (\ref{rg}) that $S$ remains unitary under the RG flow;
it also remains symmetric if it begins with a symmetric form. The fixed
points of Eq. (\ref{rg}) are given by the condition $SM^\dagger = MS^\dagger$,
i.e., $SM^\dagger$ must be Hermitian.

We can study the linear 
stability of a fixed point by deviating slightly from it, and seeing how the 
deviation grows to first order under the RG flow. Let us denote a fixed
point by the matrix $S_0$ and a deviation by $\epsilon S_1$, where $\epsilon$
is a small real parameter and $S_1$ is a matrix; we require that the matrix 
$S = S_0 + \epsilon S_1$ is unitary up to order $\epsilon$. (We can think of 
$S_1$ as defining the ``direction" of the deviation). We substitute $S$
in Eq. (\ref{rg}) and then demand that $S_1$ should take such a form that
the RG equation reduces to
\beq
\frac{d\epsilon}{dl} ~=~ \mu \epsilon ~,
\eeq
where $\mu$ is a real number. We then call the direction $S_1$ stable, 
unstable and marginal (to first order) if $\mu <0, >0$ and $0$ 
respectively. All fixed points have at least one exactly
marginal direction which corresponds to multiplying the matrix $S_0$ 
by a phase; clearly this leaves Eq. (\ref{rg}) invariant.

In this paper, we will be concerned with the RG flow of $S$-matrices which 
are 2, 3 and 4 dimensional. For convenience, we will assume certain symmetries
in each of these cases. It is useful to discuss these symmetries here,
and how they lead to some simplifications for the RG flows.

We first consider a two-wire system in which there is complete symmetry 
between the wires which we will label as 1 and 3. 
Namely, the interaction parameters are equal, $\alpha_1 = \alpha_3 =\alpha$, 
and the scattering matrix has the form
\bea
S_{2D} ~=~ \left( \begin{array}{cc} a & b \\
b & a \end{array} \right) ~.
\eea
Unitarity implies that we can parametrize $a$ and $b$ as
\bea
a &=& - ~\frac{i \lambda e^{i \theta}}{1 ~+~ i \lambda} ~, \non \\
b &=& \frac{e^{i \theta}}{1 ~+~ i \lambda} ~,
\label{smat2}
\eea
where $\lambda$ and $\theta$ are real.
Eq. (\ref{rg}) then leads to the following differential equations
\bea
\frac{d\lambda}{dl} &=& \alpha ~\lambda ~, \non \\
\frac{d\theta}{dl} &=& \frac{\alpha \lambda}{1 ~+~ \lambda^2} ~.
\label{rg2}
\eea
The reflection and transmission probabilities $|a|^2$ and $|b|^2$ only depend 
on $\lambda$. For $\alpha > 0$, we see that there is an unstable fixed
point at $\lambda =0$, and a stable fixed point at $\lambda = \infty$. If 
$\lambda$ is not zero initially (i.e., at the microscopic length scale $d$), 
then it flows to infinity at long distances. Hence $b$ goes to zero as
\beq
t ~\sim ~e^{-\alpha l} ~\sim ~L^{-\alpha} ~,
\eeq
$a$ approaches 1, and the two wires effectively get cut off from each other. 
This is in agreement with the results obtained using bosonization 
\cite{gogolin}.

Next, we will consider the $3 \times 3$ case. Here we will assume that there
is complete symmetry between two of the wires, say, 1 and 2, and that the
$S$-matrix is real. Namely, $\alpha_1 = \alpha_2$, and $S$ takes the form
\bea
S_{3D} ~=~ \left( \begin{array}{ccc} r' & t' & t \\
t' & r' & t \\
t & t & r \end{array} \right) ~,
\label{smat31}
\eea
where $r'$, $t'$ and $t$ are real parameters which, by unitarity, satisfy
\bea
t' &=& 1 ~+~ r' ~, \non \\
r &=& - ~1 ~-~ 2 r' ~, \non \\
t &=& \sqrt{(-2r')~(1+r')} ~,
\label{smat32}
\eea
and $-1 \le r' \le 0$. The RG equations in Eq. (\ref{rg}) can be written purely
in terms of the parameter $r'$ as
\beq
\frac{dr'}{dl} ~=~ - ~r' ~(1+r') ~[~ \alpha_1 r' ~+~ \alpha_3 (1 + 2r') ~]~ .
\label{rg3}
\eeq
If $\alpha_1 , \alpha_3 > 0$, we have stable fixed points at $r' =0$ (where
there is perfect transmission between wires 1 and 2, and wire 3 is cut off 
from the other two wires) and $-1$ (where all three wires are cut off from 
each other). There is also an unstable fixed point at 
\beq
r' ~=~ - ~\frac{\alpha_3}{\alpha_1 + 2 \alpha_3} ~.
\label{unstable}
\eeq
If $r'$ starts with a value which is greater than (or less than) this, then 
it flows to the value 0 (or $-1$) at large distances. [We should point out
that the fixed point in which wires 1 and 2 transmit perfectly into each other 
and wire 3 is cut off is stable only within the restricted space described 
by Eqs. (\ref{smat31}-\ref{smat32}). If we take a general unitary
matrix $S_{3D}$, then this is not a completely stable fixed point. The
only stable fixed point in the general case is the one in which all
three wires are cut off from each other \cite{lal2}.]

Finally, let us consider the $4 \times 4$ case. Here we will be interested
in a situation in which there is complete symmetry between wires 1 and 2,
and and between wires 3 and 4; further, we will take the values of $\alpha_i$
in all the wires to be equal to $\alpha$. The $S$-matrix takes the form
\bea
S_{4D} ~=~ \left( \begin{array}{cccc} a & b & c & c \\
b & a & c & c \\
c & c & a & b \\
c & c & b & a \\ \end{array} \right) ~,
\label{smat4}
\eea
where $a$, $b$ and $c$ are all complex. Unitarity implies that these 
parameters can be written in terms of three independent real variables. 
There does not seem to be a convenient parametrization in terms of 
which the RG equations take a simple form. We therefore have to study the 
RG equations in Eq. (\ref{rg}) numerically; the results will be described in 
Sec. VI. However, the fixed points of the RG equations and their linear 
stabilities can be found analytically. There are three kinds of fixed points.

\noi
(i) $|a|=1$, and $b=c=0$. This corresponds to all the wires being cut off 
from each other. This fixed point is stable in two directions, and is exactly 
marginal in one direction (corresponding to a phase rotation of $a$).

\noi
(ii) $|b|=1$, and $a=c=0$. This corresponds to perfect transmission
between wires 1 and 2, and between wires 3 and 4, but no transmission between
any other pair of wires. This fixed point is unstable in one direction (where
it flows to the fixed point described in (i)), and marginal in two directions.
One of these marginal directions turns out to be unstable at a higher order,
and the RG flow eventually takes it to the third fixed point described below. 
The other marginal direction corresponds to a phase rotation of $b$.

\noi
(iii) $|a|=1/2$, $b = - a$, and $c = \pm a$. This is a special point which 
corresponds to the maximum possible transmission with complete symmetry 
between all the four wires. This fixed point is unstable in one direction
(where it flows to the fixed point in (i)), stable in a second direction 
(where it flows in from the fixed point in (ii)), and exactly marginal in 
the third direction
(corresponding to a simultaneous phase rotation of $a$, $b$ and $c$).
The fact that (iii) is stable in one direction and unstable in another, means 
that an interesting cross-over can occur as a result of the RG flow. Namely, 
one can begin near (ii), approach (iii) for a while, and eventually go to 
(i). As a result, $|c|$ can first increase and then decrease as we go to long 
distances. This will be discussed in more detail in Sec. VI (see Fig. 15).

We see that the only completely stable fixed point is given by (i). As we 
approach this point at large distances, $b$ and $c$ go to zero as 
\beq
b, ~c ~\sim ~L^{-\alpha} ~,
\label{rg4}
\eeq
while the ratio $b/c$ approaches a constant.

Let us now consider the three systems shown in Figs. 1-3. 
In all the systems, there are four length scales of interest. 
First, there is the microscopic length scale $d$ which will
be assumed to be much smaller than all the other length scales. Then there 
is the length $L_S$ of the various sub-systems, such as the stub in Fig. 1,
each arm of the ring in Fig. 2 and the fifth wire in Fig. 3. Next, we 
have the thermal coherence length $L_T$ defined as
\beq
L_T ~=~ \frac{\hbar v_F}{k_B T} ~,
\label{lt}
\eeq
where $T$ is the temperature. As mentioned before, we will be interested in
three different regimes, namely, the ratio $L_T /L_S$ being much smaller than
1 (high temperature), comparable to 1 (intermediate temperature), and much
larger than 1 (low temperature). Finally, we have the length $L_W$ of the long
wires, namely, wires 1 and 3 in Figs. 1 and 2, and wires
1, 2, 3 and 4 in Fig. 3. We will assume that $L_W$ is much longer than both 
$L_S$ and $L_T$. The long wires will be assumed to be connected to some
reservoirs beyond the distance $L_W$. However, we will not need to consider the
reservoirs explicitly in this paper, and the length scale $L_W$ will not
enter anywhere in our calculations.

The interpretation of $L_T$ is that it is the distance beyond which the phase
of an electron wave packet becomes uncorrelated. This can be understood as 
follows. We recall that if the bias which drives the current through a QW
system is infinitesimal, then the electrons coming into the QW from 
the reservoirs have an energy $E_F = \hbar^2 k_F^2 /(2m)$, where $E_F$ is the
Fermi energy in the reservoirs. At a temperature $T$, the electron energy will
typically be smeared out by an amount of the order of $k_B T$. The uncertainty
in energy is therefore given by $k_B T = \Delta E = \hbar v_F \Delta k_F$,
where we have used Eq. (\ref{vf}). Hence, $\Delta k_F = k_B T /(\hbar v_F)
= 1/L_T$. If an electron with one particular wave number $k_F$ travels a 
distance $L$, the phase of its wave function changes by the amount $k_F L$. 
Hence, the phases of different electrons whose wave numbers vary by an 
amount $\Delta k_F$ will differ by about $\pi$ (and can therefore be
considered to be uncorrelated) if they travel a distance of about $\pi L_T$. 
Hence $L_T$ (or $\pi L_T$) can be thought of as the phase relaxation length 
of a wave packet \cite{buttiker3}. 

We can now discuss in broad terms the RG procedure that we will use for the 
various systems. In each case, we will begin at the microscopic length scale 
$d$ with certain values for the entries of the $3 \times 3$ $S$-matrices at the
various junctions. We will use Eq. (\ref{rg}) to evolve all the $S$-matrices. 
We will follow this evolution till we get to the length scale $L_S$ or $L_T$,
whichever is {\it shorter}. Two possibilities arise at this stage. 

\noi
(i) If $L_T$ is less than $L_S$, we will stop the RG flow at the length scale
$L_T$, and then calculate the transmission probabilities as discussed in Sec.
III. 

\noi
(ii) If $L_T$ is larger than $L_S$, we will stop the RG flow of the $3 \times
3$ matrices at the length scale $L_S$. Much beyond that length scale, the 
various systems shown in Figs. 1-3 look rather different since it no longer 
makes sense to consider the different junctions (and their $S$-matrices) 
separately. In particular, the stub and the ring systems look like two long 
wires joined at one point, while the four-wire system looks like four long 
wires joined at one point. Thus they all look like systems with only one 
junction as indicated in Fig. 4.
This junction is described by an effective $S$-matrix which is $2 \times 2$ 
for the stub and ring systems, and $4 \times 4$ for the four-wire system. As 
we will discuss for the different systems in Secs. IV-VI, the effective 
$S$-matrix is obtained by appropriately combining the $3 \times 3$ 
$S$-matrices at the various junctions at the length scale $L_S$; we can 
think of this process as ``integrating out" the sub-systems of length $L_S$.
Then we will continue the RG flow beyond the length scale $L_S$, but now 
with the effective $S$-matrices. This will continue till we reach the length 
scale $L_T$. At that point, we stop the RG flow and compute the transmission 
probabilities as shown in Sec. III.

[The reason for stopping the RG flow at $L_T$ in all cases is that the 
amplitudes of the various Friedel oscillations and the reflections from them 
(caused by interactions) and from the junctions are not phase coherent with 
each other beyond that length scale. Hence all these reflections will no 
longer contribute coherently to the renormalization of the scattering 
amplitudes described by the various $S$-matrices.] 

\begin{figure}[htb]
\begin{center}
\epsfig{figure=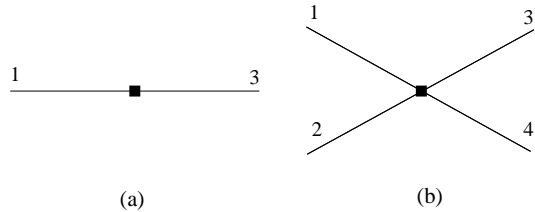,width=7cm}
\end{center}
\caption{Effective descriptions of the various systems at
low temperature, $L_T >> L_S$. The stub and ring systems effectively reduce 
to a two-wire system with a junction as in (a), while the four-wire system
reduces to a four-wire system with a junction as in (b).}
\end{figure}

To summarize, we will carry out the RG flow in one stage from the length scale
$d$ up to the length scale $L_T$, if $L_T < L_S$. If $L_T > L_S$, we will 
study the RG flow in two stages; the first stage will be with one kind of 
$S$-matrix from $d$ to $L_S$, while the second stage will be with a different 
kind of $S$-matrix from $L_S$ to $L_T$. The two kinds of $S$-matrices will be 
connected to each other at the length scale $L_S$ as discussed in Secs. IV-VI.
In all cases, when we finally stop the RG flow (after one stage or two), we 
will compute the transmission probabilities. The procedure for doing this 
will be discussed in the next section. [In reality, we expect a smooth 
cross-over from one stage of the RG procedure to the other at some length 
scale which is of the order of $L_S$. For the sake of computational 
simplicity, however, we are adopting an RG procedure which changes abruptly
exactly at $L_S$. We should also note that the RG procedures that we
will follow are really only valid for $L_T << L_S$ and $L_T >> L_S$. However,
we will assume for convenience that it is a reasonable approximation to use
the same procedures all the way up to $L_T = L_S$.]

In all the numerical results presented in Secs. IV-VI, we will take
the interaction parameter $\alpha = 0.2$ on all the wires, and the ratio
$L_S /d =10$. We will always begin the RG flow at the length scale $L_T =d$,
i.e., $L_T /L_S = 0.1$. The values of $r'$ that we will quote in the 
different figures will be the values at $L_T /L_S =0.1$.

\section{\bf Landauer-B\"uttiker Conductance}

In this section, we will discuss how to calculate the conductances of the 
various systems in the three different regimes of temperature. As mentioned
already, we assume that in each of the systems, the long wires
are eventually connected to reservoirs through some contacts. For a single
channel of spinless fermions, there is a resistance of $e^2 /h$ at the 
contacts \cite{buttiker3}. [Although the contacts can themselves scatter the 
fermions \cite{lal1}, we will ignore such effects here. We are also assuming
that the QWs are free of impurities. So the only sources of scattering in our
systems are the junctions.] 

We take the fermions in all the reservoirs to 
have the same Fermi energy $E_F$, and the net current on all the wires to be 
zero in the absence of any applied voltage on the leads. Now suppose that 
the voltage in reservoir $i$ is changed by a small amount $V_i$; here $i=1,3$ 
for the stub and ring systems, and $i=1,\cdots,4$ for the four-wire system. 
For $|V_i|$ much smaller than all the other energy scales in the problem, such
as $E_F$ and $k_B T$, the net current flowing into wire $i$ (from reservoir 
$i$) will satisfy the linear relationship \cite{buttiker3,buttiker4}
\beq
I_i ~=~ \frac{e^2}{h} ~\sum_j ~T_{ij} V_j ~,
\label{iv}
\eeq
where the $T_{ij}$ (for $i \ne j$) define the various transmission 
probabilities, and $T_{ii} +1$ denotes the reflection probability on wire $i$. 
The $T_{ij}$ satisfy certain sum rules. Current conservation implies that
\beq
\sum_i ~T_{ij} ~=~ 0
\label{sum1}
\eeq
for each value of $j$. The condition that each of the currents must be zero 
if all the $V_i$ are equal to each other implies that
\beq
\sum_j ~T_{ij} ~=~ 0
\label{sum2}
\eeq
for each value of $i$. This is equivalent to saying that changing all the 
$V_i$ by the same amount does not change any of the currents. Thus, if there 
are $N$ wires, only $N-1$ of the voltages are independent variables as far as 
the currents are concerned.

We can compute any of the conductances of the system 
if we know the values of all the $T_{ij}$ in Eq. (\ref{iv}).
One way to define a conductance is as follows \cite{buttiker3,buttiker4}.
We consider two of the long wires, say, $i$ and $j$; we call these the 
current probes, and the currents at these two wires satisfy $I_i = - I_j$. 
On all the other wires, we impose the voltage probe condition $I_m =0$; this 
imposes $N-2$ conditions on the voltages. These 
conditions imply that there is only one independent variable left 
amongst all the voltages; we can take this variable to be $V_k - V_l$,
where $k \ne l$ (in general, $k,l$ may or may not be the same as $i,j$). 
We can now define a conductance of the form
\beq
G_{ij,kl} ~=~ \frac{I_i}{V_k ~-~ V_l} ~,
\label{cond}
\eeq 

In Secs. IV and V, we will consider systems which have only two long wires 
labeled 1 and 3 (see Figs. 1 and 2). Eqs. (\ref{sum1}-\ref{sum2}) then 
imply the relations 
\beq
T_{13} ~=~ T_{31} ~=~ - ~T_{11} ~=~ - ~T_{33} ~,
\label{sym2}
\eeq
and there is only one conductance to consider, namely,
\beq
G_{13,31} ~=~ \frac{e^2}{h} ~T_{13} ~.
\eeq
In Sec. VI, we will consider a system with four long wires labeled 1, 2, 3 
and 4, with complete symmetry between wires 1 and 2 on one hand, and between 
3 and 4 on the other (see Fig. 3). In this case, we have the relations 
\bea
& & T_{ij} ~=~ T_{ji} ~, \non \\
& & T_{11} ~=~ T_{22} ~=~ T_{33} ~=~ T_{44} ~, \non \\
& & T_{12} ~=~ T_{34} ~, \non \\
& & T_{13} ~=~ T_{14} ~=~ T_{23} ~=~ T_{24} ~, \non \\
{\rm and} \quad & & T_{11} ~+~ T_{12} ~+~ 2 ~T_{13}~ ~=~ 0 ~.
\label{sym4}
\eea
There are many different conductances one can consider in this system; the 
important ones are as follows. We can take wires 1 and 2 to be the
current probes, and wires 3 and 4 to be the voltage probes. We then obtain
\beq
G_{12,21} ~=~ \frac{e^2}{h} ~(~ T_{12} ~+~ T_{13} ~) ~.
\eeq
Alternatively, we can take wires 1 and 3 to be the current probes, and 
wires 2 and 4 to be the voltage probes. We then find
\bea
G_{13,31} ~=~ \frac{4e^2}{h} ~\frac{T_{13} ~(T_{13} ~+~ T_{12})}{3T_{13} ~+~ 
T_{12}} ~, \non \\
G_{13,24} ~=~ \frac{4e^2}{h} ~\frac{T_{13} ~(T_{13} ~+~ T_{12})}{T_{13} ~-~ 
T_{12}} ~.
\eea

We now have the problem of determining the values of $T_{ij}$ in our systems.
If $L_T > L_S$, we will see in Secs. IV-VI that one can think of 
each of the systems as effectively having only one junction. In that case, 
$T_{ij}$ is related to the entries of the $S$-matrix at that junction as 
follows,
\bea
T_{ij} ~&=&~ |t_{ij}|^2 \quad {\rm for} \quad i \ne j ~, \non \\
{\rm and} \quad T_{ii} ~&=&~ |r_{ii}|^2 ~-~ 1 ~.
\label{cond2}
\eea
(Eqs. (\ref{sum1}-\ref{sum2}) then follow from the unitarity of the 
$S$-matrix). On the other hand, if $L_T < L_S$, then we have to consider 
all the junctions in the system, and the calculation of $T_{ij}$ involves 
combining the effects of several $S$-matrices in some way.

All the three systems of interest to us have two junctions. [Note that the stub
also has two junctions, i.e., a three-wire junction at the lower end $A$, and 
a one-wire junction at the upper end $B$ where we will take the $S$-matrix to 
be equal to $-1$ which corresponds to the hard wall boundary condition.] 
An electron which enters through one of the long wires has the possibility of 
bouncing back and forth many times between the two junctions. After several 
bounces, the electron can emerge from the same long wire or from a different 
long wire. From Figs. 1-3, we can see that two waves which emerge from 
the system after $n_1$ and $n_2$ bounces will have a difference in path 
lengths which is equal to $2 |n_1 - n_2| L_S$. Now we see that there are 
two regimes of temperature which will give different answers for the 
probabilities $T_{ij}$. If the thermal length $L_T$ (which is the phase 
relaxation length as argued earlier) is much smaller than $L_S$,
then the two waves will be phase incoherent if $n_1 \ne n_2$. In this case, 
the contributions of the two waves to $T_{ij}$ must be added incoherently. On 
the other hand, if $L_T$ is much larger than $2 |n_1 - n_2| L_S$, then the 
contributions of the two waves will add up coherently. In between these 
extremes is an intermediate regime in which $L_T$ is comparable to $L_S$; in 
that case, we have only partial coherence, and the two waves becomes more and 
more incoherent as $|n_1 - n_2|$ increases.

It is useful to have an expression for the $T_{ij}$ which can interpolate 
all the way from the coherent regime (low temperature) to the incoherent 
regime (high temperature). To obtain such an interpolating formula, we use 
the idea of partial coherence caused by phase randomization by a voltage 
probe which was introduced in Refs. \cite{buttiker2,mclennan}. We will first
summarize this idea, and then describe how it can be extended to our problem.

\begin{figure}[htb]
\begin{center}
\epsfig{figure=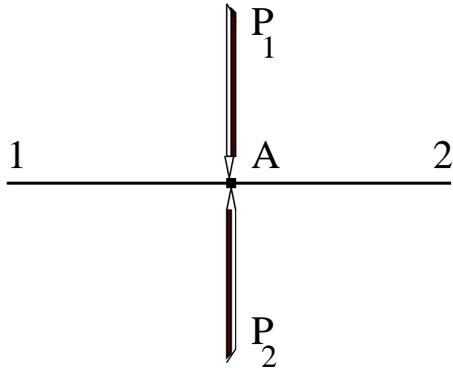,width=6cm}
\end{center}
\caption{A wire with two voltage probes $P_1$ and $P_2$ at a 
point labeled $A$. The two probes cause phase randomization of right and
left moving waves respectively.}
\end{figure}

Consider Fig. 5 in which there is a wire with two ends labeled as 1 and 2. 
At some point labeled $A$ in the middle of the wire, there are two wires
$P_1$ and $P_2$ which are voltage probes, i.e., the net outgoing currents at 
each of these wires is zero. The four-wire junction at $A$ is governed by an 
$S$-matrix of the form
\bea
S ~=~ \left( \begin{array}{cccc} 0 & \sqrt{1-p} & 0 & - \sqrt{p} \\
\sqrt{1-p} & 0 & - \sqrt{p} & 0 \\
\sqrt{p} & 0 & \sqrt{1-p} & 0 \\
0 & \sqrt{p} & 0 & \sqrt{1-p} \end{array} \right) ~,
\label{sa}
\eea
where the columns and rows carry the indices 1, 2, $P_1$ and $P_2$ in that
order, and $p$ is a real parameter which lies in the range $0 \le p \le 1$. 
(If $p=0$, the voltage probes have no effect, and phase randomization does 
not occur). A wave traveling right from end 1 can go partially out 
into $P_1$ and partially on to end 2; the part which goes out 
into $P_1$ can re-enter the wire and go on to 2. We now impose 
the phase randomization condition that the wave which
goes out into $P_1$ can re-enter the wire with an arbitrary phase change of
$\phi_1$, but it must have the same magnitude; this is necessary to ensure
the zero current condition since $P_1$ is a voltage probe. Upon solving
a problem in which there is an incoming wave of unit amplitude at end 1, and
no incoming wave at end 2, we find that there is complete
transmission of the wave across the point $A$. To be explicit, we take the
incoming and outgoing waves at $A$ to be of the forms
\bea
\psi_I ~=~ \left( \begin{array}{c} 1 \\
0 \\
e^{i\phi_1} \psi \\
0 \end{array} \right) \quad {\rm and} \quad
\psi_O ~=~ \left( \begin{array}{c} 0 \\
t \\
\psi \\
0 \end{array} \right) 
\eea
respectively. We then use the scattering matrix in Eq. (\ref{sa}) to relate
$\psi_I$ and $\psi_O$. On eliminating $\psi$, we find that 
the transmission amplitude across $A$ is given by
\beq
t ~=~ - ~e^{i\phi_1} ~\frac{1 ~-~ \sqrt{1-p} ~e^{-i \phi_1}}{1 ~-~ 
\sqrt{1-p} ~e^{i \phi_1}} ~,
\label{phase}
\eeq
so that $|t| =1$. When we calculate any physical quantity (such as a 
transmission or reflection probability), we will integrate over $\phi_1$ from 
0 to $2 \pi$. The following identity will prove to be useful,
\beq
\int_0^{2\pi} ~\frac{d\phi_1}{2\pi} ~t^n ~=~ (1-p)^{|n|/2} 
\label{tint}
\eeq
for any integer $n$. As shown below, the integration over $\phi_1$ 
reduces the coherence of a wave moving to the right from 1 to 2. Similarly, 
we can introduce a phase change of $\phi_2$ for a wave which leaves and 
re-enters the wire at the probe $P_2$; integrating over $\phi_2$ 
reduces the coherence of a wave moving to 
the left from 2 to 1. For both right and left moving waves, the degree of 
coherence depends on the value of the parameter $p$ as we will now see. 

We consider two waves which travel $n_1$
and $n_2$ times respectively through the point $A$ from left to right.
Let us suppose that their amplitudes are $a_1$ and $a_2$ respectively
in the absence of phase randomization, i.e., for $p=0$. In the presence 
of phase randomization, their amplitudes will be $a_1 t^{n_1}$ and 
$a_2 t^{n_2}$ respectively, where $t$ is given in Eq. (\ref{phase}). If these
two waves contribute to a transmission probability $T$, the cross-term 
coming from their interference will be given by $a_1^* a_2 t^{n_2 - n_1} + 
a_2^* a_1 t^{n_1 - n_2}$. We now integrate this expression over the variable
$\phi_1$. Using Eq. (\ref{tint}), we find that
\bea
& & \int_0^{2\pi} ~\frac{d\phi_1}{2\pi} ~[~ a_1^* a_2 ~t^{n_2 - n_1} ~+~ 
a_2^* a_1 ~t^{n_1 - n_2} ~] \non \\
& & =~ (~ a_1^* a_2 ~+~ a_2^* a_1 ~) ~(1-p)^{|n_1 - n_2|/2} ~.
\eea
We thus see that the phase randomization has the effect of multiplying the
interference of two terms by a factor which interpolates between 1 (i.e., 
complete phase coherence between the two waves) for $p=0$ and 0 (i.e., no 
phase coherence) for $p=1$. Further, the interpolating factor depends 
exponentially on $|n_1 - n_2|$ which is
proportional to the difference between the path lengths of the two waves.

Now we have to implement this idea in the systems of interest to us. We do 
this by generalizing the idea of phase randomization at a single point to 
phase randomization at a continuum of points. Let us assume that the density 
of such points in a wire is given by $\mu /L_T$, where $\mu$ is some 
dimensionless number (which is independent of any temperature or length 
scale), and that the parameter $p$ is the same at each of those points. This 
assumption for the density is motivated by our identification of $L_T$ as the 
phase relaxation length; the smaller the value of $L_T$, the more frequently 
phase relaxation should occur as an electron travels along the wire. 
Thus the number of phase relaxation points in 
an interval of length $L$ is equal to $\mu L/L_T$. Following arguments
similar to the one described above, one can show that the interference of two 
waves which pass through that length interval $n_1$ and $n_2$ times will get 
multiplied by the factor $(1-p)^{|n_1-n_2| \mu L/(2L_T)}$. If we write
$(1-p)^{\mu /2} = e^{-\nu}$, where $\nu$ is a positive dimensionless number,
we see that the interference between two waves whose path lengths differ 
by $\Delta L = |n_1 - n_2|L$ will carry a factor of 
\beq
F ~=~ \exp ~[~- ~\nu ~\frac{\Delta L}{L_T} ~] ~.
\label{factor1}
\eeq
The high temperature limit ($L_T \rightarrow 0$) corresponds to the
incoherent case in which we ignore the interference between paths
with any finite length difference; namely, we simply add up the probabilities
contributed by different paths.

In our calculations of the transmission probabilities described in Secs. IV-VI,
we will require an interpolating factor as in Eq. (\ref{factor1}) only for the
case $L_T \le L_S$. It is only in that regime that our systems have more than 
one junction which allows for a number of different paths between any pair of 
long wires. For $L_T > L_S$, each of our systems effectively simplify to a 
system which has only one junction and, therefore, only one possible path 
between any pair of long wires. Hence there will be no need to consider any 
interference terms for $L_T > L_S$. In order to make our expressions for the 
transmission probabilities match as we approach $L_T = L_S$ from 
above and below, we will use an interpolating factor $F$ which is 1 at $L_T = 
L_S$. We will therefore use a formula which is motivated by the expression in 
Eq. (\ref{factor1}) (with $\nu$ set equal to 1), but which is somewhat 
modified so that it is goes to 1 as
$L_T$ approaches $L_S$ from below. We will use the following prescription,
\bea
F ~=~ \exp ~[~ \frac{\Delta L}{L_S} ~-~ \frac{\Delta L}{L_T} ~] \quad {\rm 
for} \quad L_T \le L_S ~,
\label{factor2}
\eea
for the factor multiplying the interference of two paths differing in length
by $\Delta L$.

\section{\bf The Stub System}

We will now use the ideas developed in the previous two sections to study
the transmission probability of the stub system shown in Fig. 1. The $3 
\times 3$ scattering matrix, called $S_{3D}$, 
which governs the junction labeled $A$ will be assumed to be of the form
given in Eq. (\ref{smat31}-\ref{smat32}), with complete symmetry between the
two long wires labeled 1 and 3. At the other end of the stub labeled
$B$, we will assume a hard wall boundary condition, i.e., perfect reflection 
with a phase change of $-1$. 

We first consider the RG flow of the transmission probabilities $T_{ij}$. 
There is only one independent quantity to consider in this system, namely, 
$T_{13}$; all the others are related to it by Eq. (\ref{sym2}). 
As outlined in Sec. II, we start from the length scale $d$ and initially use 
Eq. (\ref{rg3}) to see how the various entries of $S_{3D}$ flow as functions of 
the length. If $L_T < L_S$, we follow this flow up to the length scale $L_T$ 
and then stop there. At that point, we compute $T_{13}$ as explained below.

If $L_T > L_S$, we first use Eq. (\ref{rg3}) to follow the flow up to the 
length scale $L_S$. At that point, we switch over to a $2 \times 2$ 
scattering matrix $S_{2D}$ which can be obtained from the matrix $S_{3D}$ 
that we get at that length scale from the RG calculation. The entries of 
$S_{2D} (L_S)$ and $S_{3D} (L_S)$ can be shown to be related as follows,
\bea
(S_{2D})_{11} &=& (S_{2D})_{33} ~=~ r' ~-~ \frac{t^2 \eta}{1 ~+~ r \eta} ~, 
\non \\
(S_{2D})_{13} &=& (S_{2D})_{31} ~=~ t' ~-~ \frac{t^2 \eta}{1 ~+~ r \eta} ~.
\label{s23s}
\eea
where
\beq
\eta ~=~ e^{i2k_F L_S} ~.
\eeq
Eq. (\ref{s23s}) will be derived in the next paragraph.
[The phase factor $\eta$ appears because the electrons are assumed to have 
a momentum of $k_F$ in all regions; hence the wave functions have factors 
of $\exp (i k_F x)$.] Having obtained $S_{2D}$ at the length scale $L_S$, 
we then continue with the RG flow of that matrix following Eq. (\ref{rg2}). 
This flow is stopped when we reach the length scale $L_T$. At that point, 
we compute $T_{13}$ as explained below.

Eq. (\ref{s23s}) can be derived in one of two ways. The first way is to assume
an incoming wave with unit amplitude on wire 1 and no incoming wave on wire 
3, and then use the scattering matrix $S_{3D}$ at junction $A$ and the sign 
change at $B$. The second way, which is more instructive for us and is also 
easier, is to sum over all the paths that an electron can take. For instance, 
if we consider the different paths which go through the stub, we see that they
are characterized by an integer $n=0,1,\cdots$ which is the number of times a 
path goes up and down the stub. The length of a path which goes from a point 
just to the left of $A$ to itself after going up and down the stub $n$ times 
is given by $2nL_S$. Summing over all such paths leads to the expression
\bea
(S_{2D})_{11} &=& r' - t \eta t + t \eta r \eta t - t \eta r \eta r \eta t + 
\cdots \non \\
&=& r' ~-~ \frac{t^2 \eta}{1 ~+~ r \eta} ~,
\label{cohs}
\eea
which is the first equation in Eq. (\ref{s23s}). Similarly, we can derive the
second equation in Eq. (\ref{s23s}) by summing over all the paths which go from
a point just to the left of $A$ to a point just to the right of $A$

Let us now calculate the transmission probability $T_{13}$. 
If $L_T \le L_S$, we have to use $S_{3D}$ to compute an expression 
for $T_{13}$ with an interpolating factor $F$ as in Eq. (\ref{factor2}). This
is easy to do since we have already found the sum over all the paths
as in Eq. (\ref{cohs}). According to the phase randomization idea discussed
in Sec. III, the interference between two paths 
characterized by integers $n_1$ and $n_2$ must be multiplied by a factor
$F = f^{|n_1 - n_2|}$, where 
\beq
f ~=~ \exp ~[~ 2 ~(~ 1 ~-~ \frac{L_S}{L_T} ~)~ ] ~.
\label{factor3}
\eeq
This follows from the prescription Eq. (\ref{factor2}) since the 
difference in path lengths is given by $\Delta L = 2 |n_1 - n_2| L_S$.
On summing up all the terms with the appropriate factors of $f$, we find that
\bea
T_{13} &=& t'^2 ~+~ \frac{t^2}{2} \non \\
& & -~ t^2 ~(~ t' ~+~ \frac{r'}{2} ~)~ [ ~\frac{\eta f}{1 + r \eta f} + 
\frac{\eta^* f}{1 + r \eta^* f} ~] ~,
\label{inters}
\eea
where we have used some of the relations in Eq. (\ref{smat32}). Eq. 
(\ref{inters}) is the desired interpolating expression for $T_{12}$. If we 
set $f=0$ (as we must do for $L_T << L_S$), we get the incoherent expression
\beq
T_{13} ~=~ t'^2 ~+~ \frac{t^2}{2} ~=~ 1 ~+~ r' ~,
\label{incohs}
\eeq
which is independent of $\eta$. On the other hand, if we set $f =1$ (as we 
must do at $L_T = L_S$), we get the coherent expression
\beq
T_{13} ~=~ |~ t' ~-~ \frac{t^2 \eta}{1 ~+~ r \eta} ~|^2
\label{coht12}
\eeq
which is just the square of the modulus of $(S_{2D})_{13}$ given in Eq. 
(\ref{s23s}). Eq. (\ref{inters}) interpolates between the coherent and 
incoherent expressions depending on the value of $f$.

There is a way of directly obtaining the incoherent expression in 
Eq. (\ref{incohs}) {\it without} summing over paths. We will present this
derivation here; as discussed in the next section, a similar derivation will 
work for the ring system where it is difficult to classify the different 
paths in a convenient way and therefore to sum over them. The idea is to add 
probabilities (intensities)
rather than amplitudes. Consider a situation with the following 
kinds of waves: a wave of unit intensity which comes into the system from wire
1, a wave of intensity $i_r$ which is reflected back to wire 1, a wave of
intensity $i_t$ which is transmitted to wire 3, a wave of intensity $i_u$
which travels up along the stub 2, and a wave of intensity $i_d$ down along
the stub. [Note that the last four waves are actually made up of sums of 
several waves obtained after repeated travels up and down the stub; however,
we will not need to explicitly sum over all those paths in this way of doing 
the calculation. The summation over paths will be implicit because we are 
assuming that $i_r$, $i_t$ and $i_u$ and $i_d$ denote the {\it total} 
intensities of those four kinds of waves.] Now we use the matrix $S_{3D}$ at 
junction $A$. This gives the following relations between these intensities,
\bea
i_r &=& r'^2 ~+~ t^2 i_d ~, \non \\
i_t &=& t'^2 ~+~ t^2 i_d ~, \non \\
i_u &=& t^2 ~+~ r^2 i_d ~.
\eea
Similarly, the total reflection at the end $B$ implies that $i_u = i_d$.
Putting these relations together, we obtain 
\beq
i_t ~=~ 1 ~+~ r' ~,
\eeq
which agrees with the expression in Eq. (\ref{incohs}).

When $L_T$ becomes equal to $L_S$, $T_{12} (L_S)$ 
is equal to $|(S_{2D})_{12}|^2$ where $(S_{2D})_{12}$
is given in Eq. (\ref{s23s}). [Our formalism is designed to ensure that we 
get the same value of $T_{13}$ at $L_T = L_S$ whether we approach that point 
from the high temperature or the low temperature side.] 
Using the parametrization in Eq. (\ref{smat2}) and the 
RG equations in Eq. (\ref{rg2}), we see that $\lambda$ at a length scale 
$L_T > L_S$ is related to its value at the length scale $L_S$ as follows,
\beq
\lambda (L_T) ~=~ \Bigl( ~\frac{L_T}{L_S} ~\Bigr)^\alpha ~\lambda (L_S) ~. 
\label{lambdalt}
\eeq
Then $T_{13} (L_T)$ is given by $1/(1 + \lambda^2 (L_T))$.

In the coherent regime given by $L_T > L_S$, we observe that $T_{13}$ is equal
to 1 if $\eta = -1$ and 0 if $\eta = 1$; this follows on using Eqs. 
(\ref{smat32}) and (\ref{coht12}). We will call these resonances and 
anti-resonances 
respectively; they arise due to interference between the different paths.
For these two special values of $\eta$, $T_{13}$ remains stuck at 1 and
0 and does not flow under RG. For any other value of $\eta$, $T_{13}$
starts at a value which is less than 1; it then flows towards zero till the 
RG evolution stops at the length scale $L_T$. Note that by changing the 
electron momentum $k_F$ (this can be done by changing the gate 
voltage), we can vary the value of $\eta$ and therefore of the matrix elements
in Eq. (\ref{s23s}); we can therefore, in principle, tune the system to 
resonance. [This is in contrast to a single wire system with an impurity where
one can change the matrix elements of $S_{2D}$ only by varying the strength of
the impurity potential which may not be easy to do experimentally.] 

\vspace*{.4cm}
\begin{figure}[htb]
\begin{center}
\epsfig{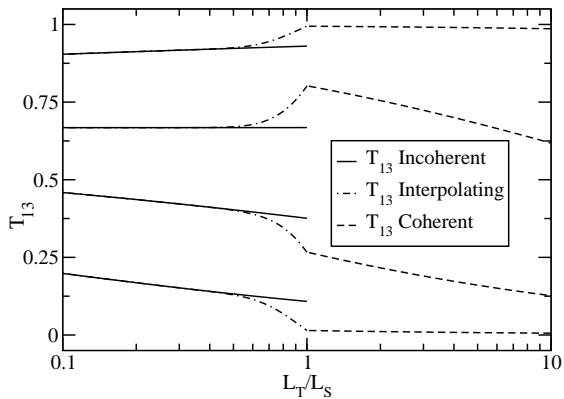}
\end{center}
\caption{$T_{13}$ for the stub system
as a function of $L_T/L_S$ for $\alpha = 0.2$,
$L_S /d = 10$, $\eta = e^{i\pi /2}$, and different values of $r'$. The four 
sets of curves are for $r' = -0.10, -0.33, -0.54$ and $-0.80$ from top to 
bottom.}
\end{figure}

\begin{figure}[htb]
\begin{center}
\epsfig{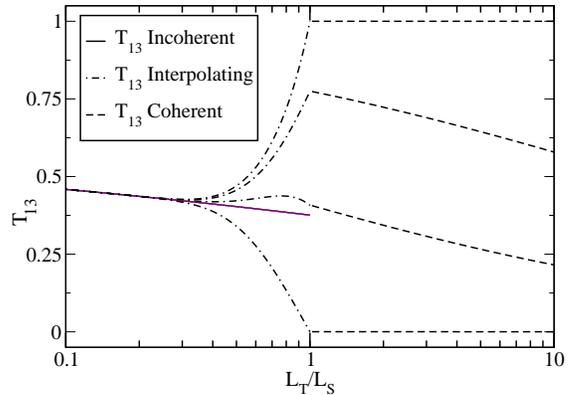}
\end{center}
\caption{$T_{13}$ as a function of $L_T/L_S$ for $\alpha = 0.2$,
$L_S /d = 10$, $r' = -0.54$, and different values of $\eta$. The four sets of
curves are for $\eta = -1, e^{i0.8 \pi}, e^{i0.6 \pi}$ and 1 from top to 
bottom.}
\end{figure}

In Figs. 6-7, we show $T_{13}$ as a function of $L_T/L_S$ for various values 
of $r'$ and $\eta$, with $\alpha = 0.2$ on all the three wires, and $L_S /d =
10$. In Fig. 6, we have considered 
four different values of $r'$. Of these values, the first one is greater than 
the unstable fixed point value of $-1/3$ given in Eq. (\ref{unstable}), the 
second is equal to $-1/3$, and the last two are less than $-1/3$.
In the incoherent regime, we see that $T_{13}$ increases in the first case, 
does not change in the second case, and decreases in the last two cases.
In Fig. 7, we show $T_{13}$ as a function of $L_T /L_S$ for four different
values of $\eta$. In the coherent regime, we see that $T_{13}$ remains
stuck at 1 and 0 for $\eta = -1$ and 1 respectively, while it decreases for
the other two cases. It is clear that modifying $\eta$ (by changing the
gate voltage) can lead to large changes in $T_{13}$ in the coherent regime.

\section{\bf The Ring System}

We now turn to the ring system shown in Fig. 2. We will assume that both
the junctions $A$ and $B$ are described by the same $3 \times 3$ scattering
matrix $S_{3D}$ given in Eqs. (\ref{smat31}-\ref{smat32}), with complete 
symmetry between the two arms of the ring labeled 2 and 4.

The RG evolution of the transmission probabilities $T_{ij}$ is studied in
the same way as for the stub system. (Once again, there is only one 
independent quantity to consider here, namely, $T_{13}$; the others are 
related to it by Eq. (\ref{sym2})). We start from the length scale $d$ and 
initially use Eq. (\ref{rg3}) to see how the various entries of the two 
matrices $S_{3D}$ flow as functions of the length. If $L_T < L_S$, we follow
this flow up to the length scale $L_T$ and then stop there. At that point, we 
compute $T_{13}$ as explained below.

If $L_T > L_S$, we first use Eq. (\ref{rg3}) to follow the flow up to the 
length scale $L_S$. At that point, we switch over to a $2 \times 2$ scattering
matrix $S_{2D}$ which can be obtained from the matrix $S_{3D}$ that we
get at that length scale from the RG calculation. For the ring system, the 
off-diagonal matrix elements of $S_{2D} (L_S)$ are related to the parameter 
$r'$ appearing in $S_{3D} (L_S)$ as follows \cite{buttiker1}.
\bea
& & (S_{2D})_{13} ~=~ (S_{2D})_{31} \non \\
& & =~ \frac{2 \cos (\Phi /2) ~\eta^{1/2} ~(1 ~-~ \eta) ~(-2r')~ (1 ~+~ r')}{[
1 ~+~ (1+2r') \eta]^2 ~-~ 2(1 ~+~ r')^2 ~(1 ~+~ \cos \Phi) ~\eta} ~, \non \\
& & 
\label{s23r}
\eea
where $\eta = e^{i2k_F L_S}$, and $\Phi$ is a dimensionless number which is 
related to the magnetic flux $\phi_B$ enclosed by the ring through the 
expression
\beq
\Phi ~=~ \frac{e\phi_B}{\hbar c} ~.
\eeq
Eq. (\ref{s23r}) will be derived in the Appendix.
Having obtained $S_{2D}$ at the length scale $L_S$, we continue with
the RG flow following Eq. (\ref{rg2}). The flow is stopped when we reach the 
length scale $L_T$. At that point, we compute $T_{13}$ as explained below.

As shown in the Appendix, Eq. (\ref{s23r}) can be obtained by assuming an 
incoming wave with unit amplitude on wire 1 and no incoming wave on wire 3, 
and then using the scattering matrices at junctions $A$ and $B$ 
\cite{buttiker1}. One might think of deriving Eq. (\ref{s23r}) by summing 
over all paths which go from wire 1 to wire 3, just as we did for the stub 
system. However, it seems very hard to enumerate the set of paths for the 
ring system in a convenient way. This is because there are two arms, and a 
path can go into either one of the two arms every time it encounters one of
the two junctions.

This difficulty in summing over paths also makes it hard to find a simple 
interpolating formula for the conductance in the regime $L_T < L_S$. To see 
this more clearly, we first note that in the {\it stub} system, two paths 
which have equal lengths must necessarily be identical to each other. Any 
point which phase randomizes waves moving in one particular direction will 
therefore occur the same number of times in the 
two paths. Hence, the interference between the two paths will not come with
any powers of either the phase factor $\eta$ or the phase randomization factor
$f$. If there are two paths of unequal lengths $2n_1 L_S$ and $2n_2 L_S$
in the stub system, then any point which phase randomizes waves moving in a
particular direction point will occur $n_1$ times in one path and $n_2$ 
times in the other path. Therefore the interference between the two paths 
will come with a factor of $\eta^{n_1-n_2} f^{|n_1 - n_2|}$. Thus, the power of
$\eta$ and the power of $f$ are always related to each other in a simple way.
(This is why a factor of $f$ always accompanies a factor of $\eta$ or $\eta^*$
in Eq. (\ref{inters})).
The situation is quite different in the ring system. Here the powers of $\eta$
and $f$ are not necessarily related to each other in any simple 
way. For instance, consider a path which enters the system through wire 1, 
goes into the arm 2 and leaves through wire 3, and a second path which enters
through wire 1, goes into the arm 4 and leaves through wire 3. These two paths
have the same length $L_S$; the interference between the two will therefore not
carry any powers of $\eta$. However, a phase randomization point which lies 
on one path will not lie on the other path. Hence, the phase randomizations
will not cancel between the two paths, and the interference between the two
paths will carry a factor of $f$. Thus, there is no general relation between 
the power of $\eta$ and the power of $f$. This makes is difficult to find
an interpolating expression for $T_{13}$.

Even though we do not have an interpolating expression for the ring system, 
we can obtain an incoherent expression for $T_{13}$ by following 
a procedure similar to the one we used for the stub system.
The idea again is to add probabilities rather than amplitudes. Consider a 
situation with the following kinds of waves: a wave of unit intensity 
which comes into the system from wire 1, a wave of intensity $i_r$ which is 
reflected back to wire 1, a wave of intensity $i_t$ which is transmitted to 
wire 3, waves of intensity $i_{2r}$ and $i_{2l}$ which travel respectively
from junction $A$ to junction $B$ and vice versa along wire 2, and waves 
of intensity $i_{4r}$ and $i_{4l}$ which travel respectively
from junction $A$ to junction $B$ and vice versa along wire 4.
[Note as before that the last six waves are actually made up of sums of 
several waves obtained after repeated bounces from the two junctions.
We do not need to explicitly sum over all these paths because we are 
assuming that $i_r, i_t, \cdots$ denote the total intensities of these 
six kinds of waves.] Now we use the matrices at junctions $A$ and $B$. 
This gives the following relations between the various intensities,
\bea
i_r &=& r^2 ~+~ t^2 ~(~ i_{2l} ~+~ i_{4l} ~) ~, \non \\
i_t &=& t^2 ~(~ i_{2r} ~+~ i_{4r} ~) ~, \non \\
i_{2r} &=& t^2 ~+~ r'^2 i_{2l} ~+~ t'^2 i_{4l} ~, \non \\
i_{4r} &=& t^2 ~+~ r'^2 i_{4l} ~+~ t'^2 i_{2l} ~, \non \\
i_{2l} &=& r'^2 i_{2r} ~+~ t'^2 i_{4r} ~, \non \\
i_{4l} &=& r'^2 i_{4r} ~+~ t'^2 i_{2r} ~.
\eea
Solving these equations and using some of the relations in Eq. (\ref{smat32}),
we find the incoherent expression for $T_{13}$ (valid for $L_T << L_S$) to be
\beq
i_t ~=~ - ~\frac{2r' ~(1 ~+~ r')}{1 ~+~ r' ~+~ r'^2} ~.
\eeq
which is independent of both $\eta$ and $\Phi$. 

At the point $L_T = L_S$, $T_{13} (L_S)$ is given by $|(S_{2D})_{13}|^2$ 
where $(S_{2D})_{13}$ is given in Eq. (\ref{s23r}). We can again use the 
parametrization in Eq. (\ref{smat2}) and the RG equation in Eq. (\ref{rg2}) to
we obtain $T_{13} (L_T) = 1/(1 + \lambda^2 (L_T))$, where 
$\lambda (L_T)$ is given in Eq. (\ref{lambdalt}).

In Fig. 8, we show $T_{13}$ as a function of $L_T /L_S$ for various
values $\eta$, with $\alpha = 0.2$ on all the wires, and $L_S /d =10$. 

\vspace*{.4cm}
\begin{figure}[htb]
\begin{center}
\epsfig{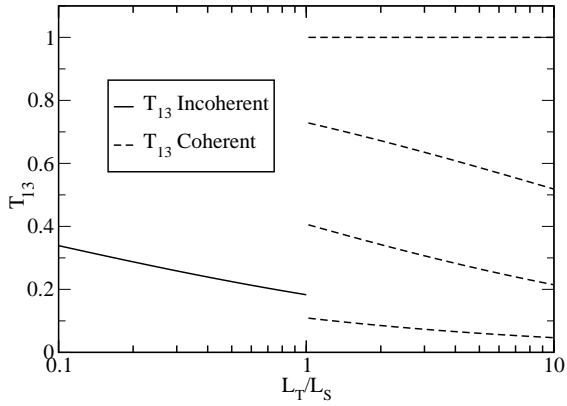}
\end{center}
\caption{$T_{13}$ for the ring system
as a function of $L_T /L_S$ for $\alpha = 0.2$, $L_S /d = 
10$, $r' = -0.18$, $\Phi =0$, and different values of $\eta$. The four coherent
curves are for $\eta = 1, e^{i 0.08 \pi}, e^{i 0.16 \pi}$ and $e^{i 0.4 \pi}$
from top to bottom. The incoherent curve is the same for all $\eta$.}
\vspace*{.4cm}
\end{figure}

\begin{figure}[htb]
\begin{center}
\epsfig{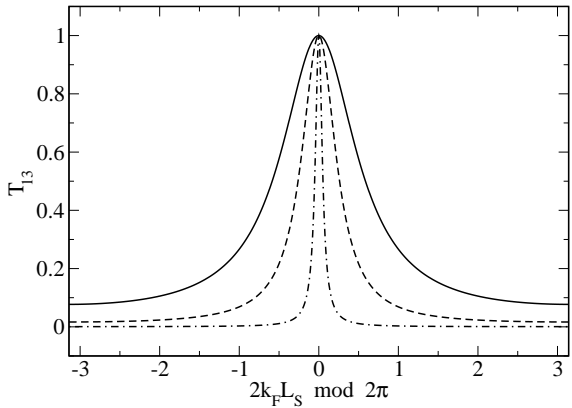}
\end{center}
\caption{$T_{13}$ as a function of $2k_F L_S$ for $\alpha = 0.2$,
$L_S /d = 10$, $r' = -0.19$, $\Phi =0$, and different values of $L_T /L_S$. 
The three sets of curves are for $L_T /L_S = 1, 50$ and $10^5$ from top to 
bottom.}
\end{figure}

\begin{figure}[htb]
\begin{center}
\epsfig{figure=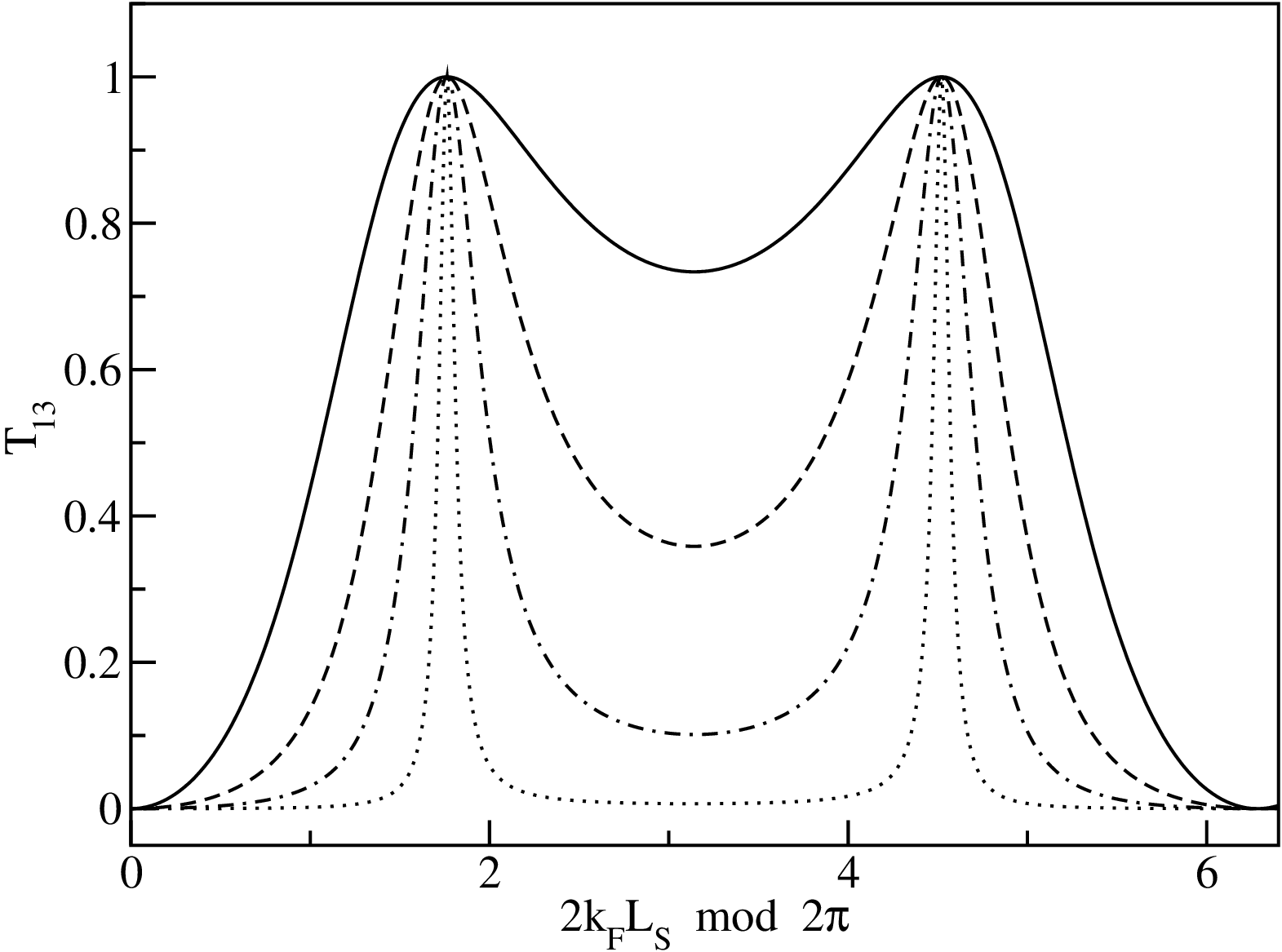,width=7.5cm}
\end{center}
\caption{$T_{13}$ as a function of $2k_F L_S$ for $\alpha = 0.2$, $L_S /d 
= 10$, $\Phi = 0.5 \pi$, and different values of $L_T /L_S$. 
The four sets of curves are for $L_T /L_S = 1, 50, 3000, $ and $10^6$
from top to bottom. Note that there is a transmission zero at $2k_F L_S = 0$ 
mod $2\pi$.}
\end{figure}

\begin{figure}[htb]
\begin{center}
\epsfig{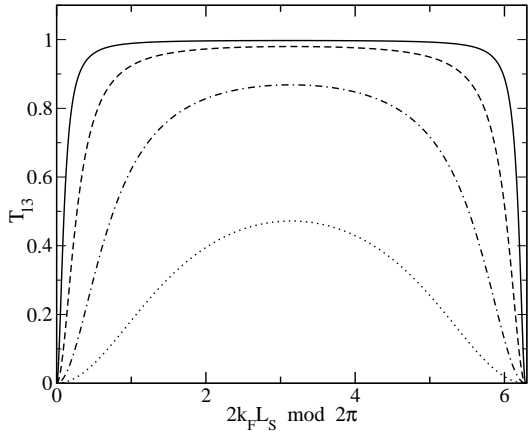}
\end{center}
\caption{$T_{13}$ as a function of $2k_F L_S$ for $\alpha = 0.2$, $L_S /d 
= 10$, $\Phi = 0$, and different values of $L_T /L_S$. The four 
sets of curves are for $L_T /L_S = 1, 150, 20000$ and $10^6$ from top to 
bottom. Note that there is a transmission zero at $2k_F L_S = 0$ mod $2\pi$.}
\end{figure}

\begin{figure}[htb]
\begin{center}
\epsfig{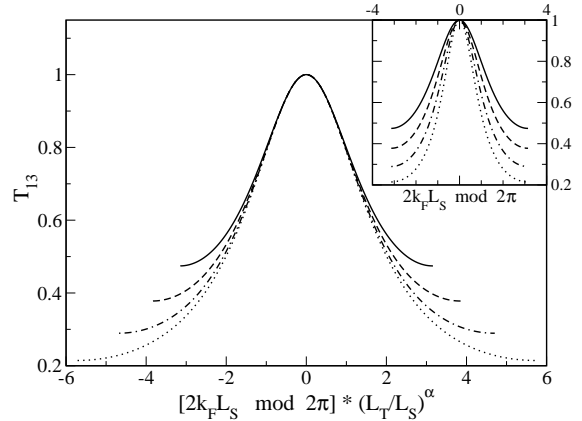}
\end{center}
\caption{$T_{13}$ as a function of the scaled variable $[2k_F L_S$ mod 
$2\pi ] ~(L_T /L_S)^\alpha$ for $\alpha = 0.2$, $L_S /d = 10$, $r' = -0.32$, 
$\Phi = 0$, and different values of $L_T /L_S$. The four sets of curves are 
for $L_T /L_S = 1, 2.7, 7.4$ and $20$ from top to bottom. The inset shows the 
same plots without scaling, i.e., $T_{13}$ as a function of $2k_F L_S$.} 
\end{figure}

In the coherent regime given by $L_T > L_S$, we can find the conditions
under which there are resonances and anti-resonances in the transmission 
through the ring, i.e., $T_{13} = 1$ and 0 respectively. (Some of these
conditions have been discussed in Ref. \cite{buttiker1}). We find that 
$T_{13} =1$ for the following values of $\eta, e^{i\Phi}$ and $r'$. 

\noi
(i) $\eta = e^{i\Phi} =1$, and $r'$ can take any value. Note that for these
values of $\eta$ and $e^{i\Phi}$, there are eigenstates of the electron
which are confined to the ring.

\noi
(ii) $\eta = e^{\pm i\Phi}$, and $r' = 0$. For this relation between 
$\eta$ and $e^{i\Phi}$, there are eigenstates of the electron on 
the ring. Further, $r' = 0$ implies $t = 0$ which means that these 
eigenstates cannot escape from the ring to the long wires. 

\noi
(iii) $e^{i\Phi} = 1$, $r' = -1/2$, and $\eta$ can take any value. Note that 
$r' = -1/2$ implies $r = 0$ which means that a wave which is coming in on wire
1 (3) suffers no reflection at junction $A$ ($B$).

\noi
[We note that both the numerator and denominator of Eq. (\ref{s23r}) vanish 
under conditions (i) and (ii); hence one has to take the limit appropriately 
to see that $T_{13} = 1$.]

Similarly, we find that $T_{13} = 0$ for the following values of $\eta$, 
$e^{i\Phi}$ and $r'$. 

\noi
(i) $\eta = 1$, and $e^{i\Phi}$ and $r'$ can take any values except 1 and -1
respectively.

\noi
(ii) $r' = 0$, and $\eta$ and $e^{i\Phi}$ can take any values except $\eta 
= e^{\pm i\Phi}$. Note that $r' = 0$ implies that $t=0$ which means that
there is no transmission between the long wires and the ring.

\noi
(iii) $r' = -1$, and $\eta$ can take any value except 1, while $e^{i\Phi}$ can
take any value. Once again, $r' = -1$ implies that $t=0$ which means that
there is no transmission between the long wires and the ring.

\noi
(iv) $e^{i\Phi} = -1$, and $(\eta, r')$ can take any values except $(1,-1)$
and $(-1,0)$.

As in the stub system, if $T_{13}$ begins with the value 1 or 0 at $L_T /L_S
=1$, it remains stuck there and does not flow under RG as we go to larger 
length scales. For any other starting value of $T_{13}$, it flows towards 
zero till the RG evolution stops at the length scale $L_T$. It is interesting 
to consider the shape of the resonance line which is a plot of $T_{13}$ 
versus the momentum $k_F$ (or, equivalently, $\eta$) at very low temperatures.
As discussed in the following paragraph, one finds that the line shape becomes
narrower with decreasing temperature, with the width at half maximum scaling 
with temperature as $T^\alpha$. Figs. 9-10 show this feature qualitatively 
for the resonances of types (i) and (ii) described above. In Fig. 10, we see 
pairs of resonances because $T_{13}$ has maxima at $2k_F L_S$ equal to $\Phi$ 
and $-\Phi$ mod $2\pi$. Fig. 11 shows the resonance of type (iii). Here 
$T_{13}$ is close to 1 for a wide range of $k_F$ (or $\eta$) at $L_T /L_S =1$;
this is consistent with the resonance condition given in (iii) above. We also 
observe anti-resonances ($T_{13} =0$) in Figs. 10 and 11 at $2k_F L_S = 0$ 
mod $2\pi$. 

Let us now discuss the resonance line shape in more
detail. Exactly at a resonance, occuring at, say, $k_F = k_{F0}$, 
$T_{13}$ is equal to 1, and it remains stuck at that value no matter how 
large $L_T$ is. We can now ask: what is the shape of the resonance line
slightly away from $k_F = k_{F0}$ ? If one deviates from $k_{F0}$ by a small 
amount $\Delta k = |k_F - k_{F0}|$ which is fixed, one finds that the 
transmission $T_{13}$ differs from 1 by an amount of order $(\Delta k)^2$. 
(An example of this is discussed below). Comparing this with the form
in Eq. (\ref{smat2}), we see that $\lambda \sim \Delta k$ at the length
scale $L_S$. Eq. (\ref{rg2}) then implies that $\lambda$ will grow as 
$\Delta k (L_T /L_S)^\alpha$ at low temperature; hence $T_{13}$ will
approach zero as $1/\lambda^2 \sim T^{2\alpha}$ at very low temperature, 
if $\Delta k$ is held fixed. On the other hand, the width of the resonance 
line at half the maximum possible value of $T_{13}$ is given by
the condition that $\Delta k (L_T /L_S)^\alpha \sim 1$, which implies that
$\Delta k \sim T^\alpha$. Thus the resonance line becomes narrower with 
decreasing temperature, with a width $\Delta k$ which vanishes as $T^\alpha$. 
To summarize, $T_{13}$ depends on the variables $\Delta k$ and $T$ through 
the combination $x= \Delta k /T^\alpha$, and $T_{13} (x) \sim 1/x^2$ as $x 
\rightarrow \infty$. (This agrees with the expression given in Ref. 
\cite{kane} for small values of $\alpha$).  As a specific 
example, let us consider the resonance of type (i). We set 
$e^{i\Phi} =1$ and take the limit $k_F \rightarrow k_{F0} = \pi n /L_S$ in 
Eq. (\ref{s23r}). We find that at the length scale $L_S$,
\beq
T_{13} ~=~ 1 ~-~ \frac{(1+2r')^2}{16r'^2 ~(1+r')^2} ~(2\Delta k L_S)^2 
\label{quad}
\eeq
up to order $(\Delta k)^2$. Eq. (\ref{rg2}) then implies that at the
length scale $L_T$,
\beq
T_{13} ~=~ \Bigl[ ~1 ~+~ \frac{(1+2r')^2}{16r'^2 ~(1+r')^2} ~
(2\Delta k L_S)^2 (\frac{L_T}{L_S})^{2 \alpha} ~\Bigr]^{-1} ~.
\eeq
Thus, if $r'$ is held fixed and $T_{13}$ is plotted against $\Delta k 
(L_T /L_S)^\alpha$, we should get the same curve for different values of 
$L_T /L_S$, provided that the quadratic approximation in Eq. (\ref{quad})
holds good. In Fig. 12, we show $T_{13}$ as a function of $2 \Delta k L_S 
(L_T /L_S)^\alpha$ for four different values of $L_T /L_S$. We see that
the curves agree well with each other down to about $T_{13} = 0.7$. For 
comparison, we have shown the same plots without scaling in the inset; 
we see that they begin disagreeing below $T_{13} = 0.95$. (We find similar 
resonance line shapes in the stub and four-wire systems, although we have not 
shown those plots in this paper).

\section{\bf The Four-wire System}

Finally, let us consider the four-wire system shown in Fig. 3. We will assume
that both the junctions $A$ and $B$ are described by the same $3 \times 3$ 
scattering matrix $S_{3D}$ given in Eqs. (\ref{smat31}-\ref{smat32}), with 
complete symmetry between the wires 1 and 2 on one side and the wires 3 and 4 
on the other side. The transmission probabilities enjoy the symmetries 
described in Eq. (\ref{sym4}).

We first consider the RG flow of the transmission probabilities $T_{ij}$.
Due to the symmetries of the system, and the relations in
Eqs. (\ref{sum1}-\ref{sum2}), we see that there are only two independent
quantities to consider, namely, $T_{12}$ and $T_{13}$. Following the formalism
in Sec. II, we start from the length scale $d$ and initially use Eq. 
(\ref{rg3}) to see how the various entries of $S_{3D}$ flow as functions of 
the length. If $L_T < L_S$, we follow this flow up to the length scale $L_T$, 
and then compute $T_{12}$ and $T_{13}$.

If $L_T > L_S$, we first use Eq. (\ref{rg3}) to follow the flow up to the 
length scale $L_S$. At that point, we switch over to a $4 \times 4$ 
scattering matrix $S_{4D}$ which can be obtained from the matrix $S_{3D}$ 
that we get at that length scale from the RG calculation. The entries of 
$S_{4D} (L_S)$ and $S_{3D} (L_S)$ can be shown to be related as follows,
\bea
(S_{4D})_{11} &=& r' ~+~ \frac{t^2 r \eta}{1 ~-~ r^2 \eta} ~, \non \\
(S_{4D})_{12} &=& t' ~+~ \frac{t^2 r \eta}{1 ~-~ r^2 \eta} ~, \non \\
(S_{4D})_{13} &=& \frac{t^2 \eta^{1/2}}{1 ~-~ r^2 \eta} ~.
\label{s43f}
\eea
where $\eta = e^{i2k_F L_S}$. (Eq. (\ref{s43f}) will be derived in the next 
paragraph). Having obtained $S_{4D}$ at the length scale $L_S$, we then 
continue
with the RG flow of that matrix using Eq. (\ref{rg}). This flow is stopped 
when we reach the length scale $L_T$, where we compute $T_{12}$ and $T_{13}$.

As in the stub system, Eq. (\ref{s43f}) can be derived in one of two ways. 
The first way is to assume an incoming wave with unit amplitude on wire 1 and 
no incoming waves on wires 2, 3 and 4, and then use the scattering matrices 
$S_{3D}$ at junctions $A$ and $B$. The second way is to sum 
over all the paths that an electron can take. As in the stub system, the
different paths going between any two of the long wires $i$ and $j$
are characterized by an integer $n=0,1,\cdots$ which is the number of times a 
path goes right and left on the central wire labeled 5. For instance, the
sum over paths which go from a point on wire 1 lying very close to the 
junction $A$ to itself gives the series
\bea
(S_{4D})_{11} &=& r' ~+~ t \eta^{1/2} r \eta^{1/2} t \non \\
& & +~ t \eta^{1/2} r \eta^{1/2} r \eta^{1/2} r \eta^{1/2} t ~+~ \cdots ~,
\eea
which agrees with the first equation in Eq. (\ref{s43f}).
Similarly, we can derive the other expressions in Eq. (\ref{s43f}).

Let us now calculate the transmission probabilities $T_{12}$ and $T_{13}$. 
If $L_T \le L_S$, we have to use $S_{3D}$ to compute expressions 
for $T_{ij}$ with an interpolating factor $f$ as in Eq. (\ref{factor3}). This
is as easy to do here as in the stub system since we know how to explicitly
sum over all the paths. The interference between the contributions of two paths
characterized by integers $n_1$ and $n_2$ must be multiplied by a factor
$f^{|n_1 - n_2|}$, where $f$ is given in Eq. (\ref{factor3}). 
On summing up all the terms with the appropriate factors of $f$, we find that
\bea
T_{12} &=& t'^2 ~+~ \frac{t^2 r^2}{2(1+r^2)} \non \\
& & + ~t^2 \Bigl( t' r + \frac{r^4}{2(1+r^2)} \Bigr) \Bigl( \frac{\eta 
f}{1 - r^2 \eta f} + \frac{\eta^* f}{1 - r^2 \eta^* f} \Bigr) , \non \\
T_{13} &=& \frac{t^2}{2(1+r^2)} + \frac{t^2 r^2}{2(1+r^2)} \Bigl( 
\frac{\eta f}{1 - r^2 \eta f} + \frac{\eta^* f}{1 - r^2 \eta^* f} \Bigr) . 
\non \\
& &
\label{interf}
\eea
These are the desired interpolating expressions for $T_{12}$ and $T_{13}$. If
we set $f=0$ (as we must do for $L_T << L_S$), we get the incoherent 
expressions
\bea
T_{12} &=& t'^2 ~+~ \frac{t^2 r^2}{2(1+r^2)} ~=~ \frac{(1+r')(2+5r' + 
4r'^2)}{2(1+2r' +2r'^2)} ~, \non \\
T_{13} &=& \frac{t^2}{2(1+r^2)} ~=~ - ~\frac{r' (1+r')}{2(1+2r' +2r'^2)} ~,
\label{incohf}
\eea
which are independent of $\eta$. On the other hand, if we set $f =1$ (as we 
must do at $L_T = L_S$), we get the coherent expressions which are given by 
the square of the modulus of the entries $(S_{4D})_{12}$ and $(S_{4D})_{13}$ 
in Eq. (\ref{s43f}). 

As in the stub system, there is a way of directly obtaining the incoherent 
expression in Eq. (\ref{incohf}) without summing over paths, by adding 
probabilities rather than amplitudes. We consider a situation with the 
following kinds of waves: a wave of unit intensity which comes into the 
system from wire 1, waves of intensity $i_2$, $i_3$ and $i_4$ which 
go into wires 2, 3 and 4, and waves of intensity $i_r$ and $i_l$ which
travel right and left respectively on wire 5. We then use the matrices
$S_{3D}$ at junctions $A$ and $B$ to write down the relations
between all these intensities, and then solve for $i_2$ and $i_3$. 
This reproduces the results in Eq. (\ref{incohf}).

%\begin{figure}[htb]
%\begin{center}
%\epsfig{figure=Dy1.eps,width=7.5cm}
%\end{center}
%\caption{$T_{12}$ as a function of $L_T/L_S$ for $\alpha = 0.2$, $L_S /d = 
%10$, and $\eta =1$. The four sets of curves (incoherent and interpolating) 
%are for $r' = -0.07, -0.17, -0.28$ and $-0.43$ from top to bottom. The 
%coherent curve remains stuck at $1/4$ for all values of $r'$ since $\eta =1$.}
%\end{figure}

\vspace*{.4cm}
\begin{figure}[htb]
\begin{center}
\epsfig{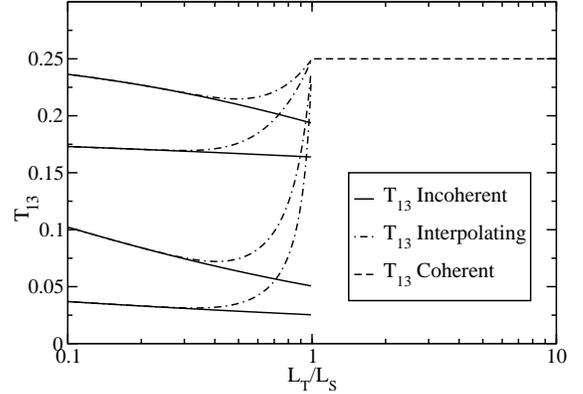}
\end{center}
\caption{$T_{13}$ for the four-wire
system as a function of $L_T/L_S$ for $\alpha = 0.2$, $L_S /d = 
10$, and $\eta =1$. The four sets of curves (incoherent and interpolating) 
are for $r' = -0.43, -0.28, -0.17$ and $-0.07$ from top to bottom. The coherent
curve remains stuck at $1/4$ for all values of $r'$ since $\eta =1$.}
\end{figure}

%\begin{figure}[htb]
%\begin{center}
%\epsfig{figure=Dy3.eps,width=7.5cm}
%\end{center}
%\caption{$T_{12}$ as a function of $L_T/L_S$ for $\alpha = 0.2$, $L_S /d = 
%10$, and $r'=-0.28$. The four sets of curves (interpolating and coherent) are 
%for $\eta = e^{i 0.9 \pi}, e^{i 0.6 \pi}, e^{i 0.3 \pi}$ and 1 from top to 
%bottom. The incoherent curve is independent of $\eta$.}
%\end{figure}

\begin{figure}[htb]
\begin{center}
\epsfig{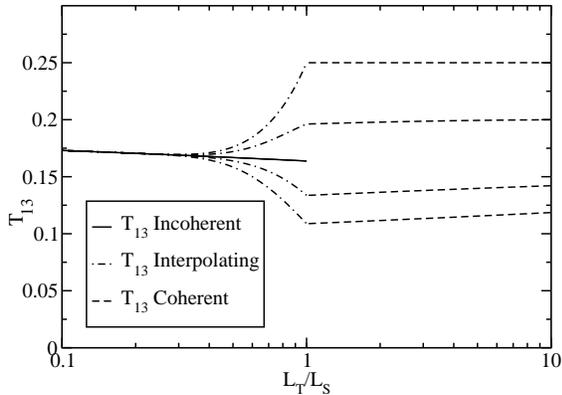}
\end{center}
\caption{$T_{13}$ as a function of $L_T/L_S$ for $\alpha = 0.2$, $L_S /d = 
10$, and $r'=-0.28$. The four sets of curves (interpolating and coherent) are
for $\eta = 1, e^{i 0.3 \pi}, e^{i 0.6 \pi}$ and $e^{i 0.9 \pi}$ from top to 
bottom. The incoherent curve is independent of $\eta$.}
\vspace*{.4cm}
\end{figure}

\begin{figure}[htb]
\begin{center}
\epsfig{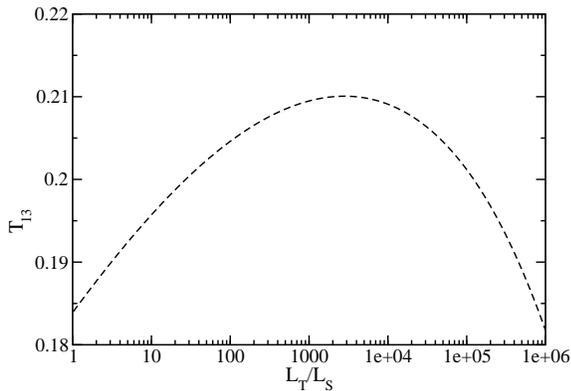}
\end{center}
\caption{$T_{13}$ as a function of $L_T /L_S$ for $\alpha = 0.2$, $L_S /d = 
10$, $r' = -0.33$ and $\eta = e^{i0.52 \pi}$. $T_{13}$ first increases and 
then decreases at very low temperatures.}
\end{figure}

If $L_T > L_S$, $T_{12}$ and $T_{13}$ are equal to $|(S_{4D})_{12}|^2$ and
$|(S_{4D})_{13}|^2$, where $(S_{4D})_{12}$ and $(S_{4D})_{13}$ are
given in Eq. (\ref{s43f}). In this regime, the RG flow has to be carried out
numerically for the reasons explained after Eq. (\ref{smat4}). In general, we 
find that at long distances, $b$ and $c$ flow to zero as indicated in 
Eq. (\ref{rg4}); hence $T_{12}$ and $T_{13}$ go to zero as $L^{-2\alpha}$.

In the coherent regime given by $L_T > L_S$, we observe that $T_{12}$ and
$T_{13}$ are both equal to $1/4$ if either $\eta = 1$ or $r' = -1/2$. 
We may call these resonances since the maximum possible value of $T_{13}$ 
which is allowed by the form of the matrix in Eq. (\ref{smat4}) is $1/4$.
If $\eta =1$, $T_{13}$ remains stuck at $1/4$ and does not flow under RG. 
This can be seen in Fig. 13 where we show $T_{13}$ as a function of $L_T/L_S$ 
for various values of $r'$. For any other value of $\eta$, $T_{13}$ flows 
till the RG evolution stops at the length scale $L_T$. (As discussed below,
$T_{13}$ can sometimes increase before eventually decreasing towards
zero at very low temperatures). As in the 
stub system, we can vary the value of $\eta$ and therefore tune the system to 
resonance by changing the electron momentum $k_F$.

In Fig. 14, we show $T_{13}$ as a function of $L_T/L_S$ for various values of 
$\eta$. In Fig. 15, we show the cross-over behavior of $T_{13}$ mentioned in 
Sec. II. In the coherent regime, for certain ranges of values of $r'$ and 
$\eta$, $T_{13}$ first increases and then decreases at very low temperatures.

\section{\bf Discussion}

In this work, we have derived the RG equations and the transmission
probabilities (and conductances) for three systems of experimental interest. 
The RG flows and the consequent power-laws in the temperature and length 
dependences of the conductances are purely a result of the interactions in 
the wires; there is no RG flow if the interaction parameters $\alpha_i$ are 
all zero. A peculiarity of our RG formalism is that it has two stages 
which work in the regimes of high and low temperature respectively. 
We abruptly switch between the two stages when we cross the point $L_T/L_S 
=1$. It would be useful to develop an interpolating formalism for the RG flow
which can vary smoothly across the intermediate range of temperature. In
our way of deriving the RG equations, this may require an analysis of the
way in which Friedel oscillations from two junctions interfere with each other.

Our results should be applicable to the systems mentioned earlier such as 
multi-arm quantum wires \cite{timp,shepard}, various kinds of carbon 
nanotubes \cite{papa,kim1}, and systems with other kinds of geometry
\cite{petrashov,casse,debray}. While some of the early experiments focussed on 
electronic transport in the presence of an external magnetic field and the
effects of geometry, measuring the various conductances at different 
temperatures (and, if possible, different wire lengths) should reveal the 
interaction induced power-laws discussed in our work. Note that a spread in 
the phase (as discussed after Eq. (\ref{lt}) in Sec. II) and phase 
randomization (as discussed in Sec. III) are the only effects of thermal 
fluctuations that we have considered in this work. We have ignored other 
effects of finite temperature, such as momentum relaxation by inelastic 
scattering, and corrections to the Landauer-B\"uttiker conductances due to 
thermal broadening of the Fermi-Dirac distribution near the Fermi energy. An 
application of our work to experiments would require one to 
disentangle these other features before the effects of interactions can 
become visible. Typically, the temperature at which these experiments are 
done is about $0.1 - 1^o$ K, while the Fermi energy is about $10^o$ K which
is much larger \cite{casse}; hence the thermal broadening effect is
expected to be small.

One limitation of our work is that we have assumed linear relations between the
incoming and outgoing fermion fields. In principle, other interesting things 
can happen at a junction, particularly if we consider the case of spinful 
fermions and if some of the wires are superconducting rather than metallic. 
For instance, there may be Andreev reflection in which an electron striking the
junction from one wire is reflected back as a hole while two electrons are 
transmitted into some of the other wires \cite{petrashov,nayak}. It would be 
interesting to study these phenomena using the techniques developed in this 
paper.

\vskip .6 true cm
\centerline{\bf ACKNOWLEDGMENTS}
\vskip .6 true cm

D.S. acknowledges financial support from a Homi Bhabha Fellowship and the 
Council of Scientific and Industrial Research, India through Grant No. 
03(0911)/00/EMR-II.

\newpage

\vskip .6 true cm
\centerline{\bf APPENDIX}
\vskip .6 true cm

We will derive Eq. (\ref{s23r}) here. We consider a situation with the 
following kinds of waves: an incoming wave on wire 1 whose amplitude is
unity just to the left of junction $A$, an outgoing wave on wire 1 whose 
amplitude is $\psi_{1,{\rm out}}$ just to the left of junction $A$, 
an outgoing wave on wire 3 whose amplitude is $\psi_{3,{\rm out}}$ just to 
the right of junction $B$, waves on wires 2 and 4 which are 
incoming at junction $A$ and have amplitudes $\psi_{2,{\rm in}}^A$ and
$\psi_{4,{\rm in}}^A$ just to the right of $A$, waves on wires 2 and 4 
which are outgoing at junction $A$ and have amplitudes $\psi_{2,{\rm out}}^A$
and $\psi_{4,{\rm out}}^A$ just to the right of $A$, waves on wires 2 and 4 
which are incoming at junction $B$ and have amplitudes $\psi_{2,{\rm in}}^B$ 
and $\psi_{4,{\rm in}}^B$ just to the left of $B$, and waves on wires 2 and 4 
which are outgoing at junction $B$ and have amplitudes $\psi_{2,{\rm out}}^B$
and $\psi_{4,{\rm out}}^B$ just to the left of $B$. Our aim is to find an
expression for the transmitted amplitude $\psi_{3,{\rm out}}$. 

The Schr\"odinger equation relates many of the amplitudes introduced above
to each other. This is because of the following features: a wave which
travels a distance $x$ picks up a phase of $e^{ik_F x}$ (we are assuming 
that all the particles have momentum $k_F$), a wave which
travels anticlockwise around the ring from junction $A$ to junction $B$
or vice versa picks up a phase of $e^{i\Phi /2}$, and a wave which
travels clockwise around the ring from junction $A$ to junction $B$
or vice versa picks up a phase of $e^{-i\Phi /2}$. This gives us the
following relations:
\bea
\psi_{2,{\rm in}}^B &=& e^{ik_F L_S - i \Phi /2} ~\psi_{2,{\rm out}}^A ~, 
\non \\
\psi_{2,{\rm in}}^A &=& e^{ik_F L_S + i \Phi /2} ~\psi_{2,{\rm out}}^B ~, 
\non \\
\psi_{4,{\rm in}}^B &=& e^{ik_F L_S + i \Phi /2} ~\psi_{4,{\rm out}}^A ~, 
\non \\
\psi_{4,{\rm in}}^A &=& e^{ik_F L_S - i \Phi /2} ~\psi_{4,{\rm out}}^B ~.
\label{app1}
\eea
Now we use the form of the scattering matrices in Eq. (\ref{smat31}) at 
the two junctions. At junction $A$, we have
\bea
\psi_{1,{\rm out}} &=& r ~+~t ~(~ \psi_{2,{\rm in}}^A ~+~ 
\psi_{4,{\rm in}}^A ~) ~, \non \\
\psi_{2,{\rm out}}^A &=& t ~+~ r' \psi_{2,{\rm in}}^A ~+~ 
t' \psi_{4,{\rm in}}^A ~, \non \\
\psi_{4,{\rm out}}^A &=& t ~+~ r' \psi_{4,{\rm in}}^A ~+~ 
t' \psi_{2,{\rm in}}^A ~. 
\label{app2}
\eea
At junction $B$, we have
\bea
\psi_{3,{\rm out}} &=& t ~(~ \psi_{2,{\rm in}}^B ~+~ \psi_{4,{\rm in}}^B ~) ~,
\non \\
\psi_{2,{\rm out}}^B &=& r' \psi_{2,{\rm in}}^B ~+~ t' \psi_{4,{\rm in}}^B ~,
\non \\
\psi_{4,{\rm out}}^B &=& r' \psi_{4,{\rm in}}^B ~+~ t' \psi_{2,{\rm in}}^B ~. 
\label{app3}
\eea
Using Eqs. (\ref{app1}-\ref{app3}), we obtain the expression for 
$(S_{2D})_{31} =\psi_{3,{\rm out}}$ given in Eq. (\ref{s23r}).


\begin{thebibliography}{99}

\bibitem{timp} G. Timp, R. E. Behringer, E. H. Westerwick, and J. E. 
Cunningham, in {\it Quantum Coherence in Mesoscopic Systems}, edited by 
B. Kramer (Plenum Press, 1991) pages 113-151 and references therein; 
C. B. J. Ford, S. Washburn, M. B\"uttiker, C. M. Knoedler, and J. M. Hong, 
Phys. Rev. Lett. {\bf 62}, 2724 (1989).

\bibitem{shepard} K. L. Shepard, M. L. Roukes, and B. P. Van der Gaag, 
Phys. Rev. Lett. {\bf 68}, 2660 (1992).
 
\bibitem{papa} C. Papadopoulos, A. Rakitin, J. Li, A. S. Vedeneev, and 
J. M. Xu, Phys. Rev. Lett. {\bf 85}, 3476 (2000).

\bibitem{kim1} J. Kim, K. Kang, J.-O. Lee, K.-H. Yoo, J.-R. Kim, J. W. 
Park, H. M. So, and J.-J. Kim, J. Phys. Soc. Jpn. {\bf 70}, 1464 (2001).

\bibitem{petrashov} V. T. Petrashov, V. N. Antonov, P. Delsing, and R. Claeson,
Phys. Rev. Lett. {\bf 70} (1993) 347.

\bibitem{casse} M. Casse, Z. D. Kvon, G. M. Gusev, E. B. Olshanetskii, L. V. 
Litvin, A. V. Plotnikov, D. K. Maude, and J. C. Portal, Phys. Rev. B {\bf 62},
2624 (2000). 

\bibitem{debray} P. Debray, O. E. Raichev, P. Vasilopoulos, M. Rahman, R. 
Perrin, and W. C. Mitchell, Phys. Rev. B {\bf 61}, 10950 (2000); R. Akis, P. 
Vasilopoulos, and P. Debray, {\it ibid.} {\bf 56}, 9594 (1997); R. Akis, P. 
Vasilopoulos, and P. Debray, {\it ibid.} {\bf 52}, 2805 (1995).

\bibitem{buttiker1} M. B\"uttiker, Y. Imry, and M. Ya. Azbel, Phys. Rev. A 
{\bf 30}, 1982 (1984).

\bibitem{jayannavar} T. P. Pareek and A. M. Jayannavar, Phys. Rev. B {\bf 54},
6376 (1996); T. P. Pareek, P. Singha Deo, and A. M. Jayannavar, {\it ibid.}
{\bf 52}, 14657 (1995); P. Singha Deo and A. M. Jayannavar, {\it ibid.} {\bf 
50}, 11629 (1994).

\bibitem{shi} Y. Shi and H. Chen, Phys. Rev. B {\bf 60}, 10949 (1999).

\bibitem{deo} P. Singha Deo and M. V. Moskalets, Phys. Rev. B {\bf 61}, 10559
(2000); M. V. Moskalets and P. Singha Deo, {\it ibid.} {\bf 62}, 6920 (2000).

\bibitem{kim2} T.-S. Kim, S. Y. Cho, C. K. Kim, and C.-M. Ryu, Phys. Rev. B 
{\bf 65}, 245307 (2002).

\bibitem{kane} C. L. Kane and M. P. A. Fisher, Phys. Rev. B {\bf 46}, 15233
(1992).

\bibitem{safi} I. Safi and H. J. Schulz, Phys. Rev. B {\bf 52}, 17040 (1995); 
D. L. Maslov and M. Stone, {\it ibid} {\bf 52}, 5539 (1995); V. V. Ponomarenko,
{\it ibid.} {\bf 52}, 8666 (1995).

\bibitem{maslov} D. L. Maslov, Phys. Rev. B, {\bf 52}, 14386 (1995); A. 
Furusaki and N. Nagaosa, {\it ibid} {\bf 54}, 5239 (1996); I. Safi and H. J. 
Schulz, {\it ibid} {\bf 59}, 3040 (1999).

\bibitem{lal1} S. Lal, S. Rao, and D. Sen, Phys. Rev. Lett. {\bf 87}, 026801 
(2001), Phys. Rev. B {\bf 65}, 195304 (2002), and cond-mat/0104402.

\bibitem{gogolin} A. O. Gogolin, A. A. Nersesyan, and A. M. Tsvelik, {\it
Bosonization and Strongly Correlated Systems} (Cambridge University Press,
Cambridge, 1998); S. Rao and D. Sen, in {\it Field Theories in Condensed 
Matter Physics}, edited by S. Rao (Hindustan Book Agency, New Delhi, 2001).

\bibitem{tarucha} S. Tarucha, T. Honda, and T. Saku, Sol. St. Comm. {\bf 94},
413 (1995); A. Yacoby, H. L. Stormer, N. S. Wingreen, L. N. Pfeiffer, K. W.
Baldwin, and K. W. West, Phys. Rev. Lett. {\bf 77}, 4612 (1996); C. -T. 
Liang, M. Pepper,, M. Y. Simmons, C. G. Smith, and D. A. Ritchie, Phys. Rev. 
B {\bf 61}, 9952 (2000); B. E. Kane, G. R. Facer, A. S. Dzurak, N. E. Lumpkin,
R. G. Clark, L. N. Pfeiffer, and K. W. West, App. Phys. Lett. {\bf 72}, 3506 
(1998); D. J. Reilly, G. R. Facer, A. S. Dzurak, B. E. Kane, R. G. Clark, P.
J. Stiles, J. L. O'Brien, N. E. Lumpkin, L. N. Pfeiffer, and K. W. West, 
Phys. Rev. B {\bf 63}, 121311 (2001).

\bibitem{nayak} C. Nayak, M. P. A. Fisher, A. W. W. Ludwig, and H. H. Lin,
Phys. Rev. B {\bf 59}, 15694 (1999).

\bibitem{komnik} A. Komnik and R. Egger, Phys. Rev. Lett. {\bf 80}, 2881
(1998), and Eur. Phys. J. B {\bf 19}, 271 (2001).

\bibitem{lal2} S. Lal, S. Rao, and D. Sen, Phys. Rev. B {\bf 66}, 165327 
(2002).

\bibitem{chen} S. Chen, B. Trauzettel, and R. Egger, Phys. Rev. Lett. {\bf 
89}, 226404 (2002).

\bibitem{zhu} J.-L. Zhu, X. Chen, and Y. Kawazoe, Phys. Rev. B {\bf 55}, 16300 
(1997).

\bibitem{ben} E. Ben-Jacob, F. Guinea, Z. Hermon, and A. Shnirman, Phys. Rev.
B {\bf 57}, 6612 (1998).

\bibitem{sumathi} P. Durganandini and S. Rao, Phys. Rev. B {\bf 61}, 4739 
(2000); P. Durganandini and S. Rao, {\it ibid.} {\bf 59}, 13122 (1999).

\bibitem{kim3} M. D. Kim, S. Y. Cho, C. K. Kim, and K. Nahm, Phys. Rev. B
{\bf 66}, 193308 (2002).

\bibitem{yue} D. Yue, L. I. Glazman, and K. A. Matveev, Phys. Rev. B {\bf 
49}, 1966 (1994).

\bibitem{buttiker2} M. B\"uttiker, Phys. Rev. B {\bf 33}, 3020 (1986); 
IBM J. Res. Dev. {\bf 32}, 63 (1988).

\bibitem{mclennan} M. J. McLennan, Y. Lee, and S. Datta, Phys. Rev. B {\bf 
43}, 13846 (1991). 

\bibitem{buttiker3} M. B\"uttiker, Y. Imry, and R. Landauer, Phys. Rev. B {\bf 
31}, 6207 (1985); S. Datta, {\it Electronic transport in mesoscopic systems}
(Cambridge University Press, Cambridge, 1995); {\it Transport phenomenon in
mesoscopic systems}, edited by H. Fukuyama and T. Ando (Springer Verlag,
Berlin, 1992); Y. Imry, {\it Introduction to Mesoscopic Physics} (Oxford
University Press, New York, 1997). 

\bibitem{buttiker4} M. B\"uttiker, Phys. Rev. Lett. {\bf 57}, 1761 (1986); 
IBM J. Res. Dev. {\bf 32}, 317 (1988).

\end{thebibliography}
\end{document}